\documentclass[12pt]{article}

\usepackage{graphics}
\usepackage{epsfig}
\usepackage{cite}
\input epsf

\title{Prompt charmonia production and polarization at LHC in the NRQCD with $k_T$-factorization. \\Part I: $\psi(2S)$ meson}
\author{S.P.~Baranov$^{1}$, A.V.~Lipatov$^{2,\,3}$, N.P.~Zotov$^2$}

\begin{document}

\maketitle

\begin{center}

{\it $^1$P.N.~Lebedev Physics Institute, 119991 Moscow, Russia}\\
{\it $^2$Skobeltsyn Institute of Nuclear Physics, Lomonosov Moscow State University, 119991 Moscow, Russia}\\
{\it $^3$Joint Institute for Nuclear Research, Dubna 141980, Moscow region, Russia}\\

\end{center}

\vspace{5mm}

\begin{center}

{\bf Abstract }

\end{center}

In the framework of $k_T$-factorization approach, 
the production and polarization of prompt $\psi(2S)$ mesons in $pp$ collisions 
at the LHC energies is studied. Our consideration is based on the non-relativistic
QCD formalism for bound states and off-shell amplitudes for hard partonic 
subprocesses.
The transverse momentum dependent (unintegrated) gluon densities in a proton were derived from
Ciafaloni-Catani-Fiorani-Marchesini evolution equation or, alternatively, were
chosen in accordance with Kimber-Martin-Ryskin prescription.
The non-perturbative color-octet matrix elements were first deduced from the 
fits to the latest CMS data on $\psi(2S)$ transverse momentum distributions
and then applied to describe the ATLAS and LHCb data on $\psi(2S)$ production 
and polarization at $\sqrt s = 7$~TeV.
We perform the estimation of polarization parameters $\lambda_\theta$, 
$\lambda_\phi$ and $\lambda_{\theta \phi}$ which determine $\psi(2S)$ 
spin density matrix and demonstrate that taking into account the off-shellness
of initial gluons in the color-octet contributions 
leads to unpolarized $\psi(2S)$ production at high
transverse momenta, that is in qualitative agreement with the LHC data.

\vspace{1.0cm}

\noindent
PACS number(s): 12.38.-t, 13.20.Gd, 14.40.Pq

\newpage

\section{Introduction} \indent 

The production of quarkonium states in high energy hadronic collisions
is under intense theoretical and experimental study\cite{1,2,3} since two
decades ago, when the measurements of prompt $J/\psi$ and $\Upsilon$ production 
cross sections at the Tevatron revealed a more than one order-of-magnitude 
discrepancy with theoretical expectations obtained in the framework of the 
color singlet model\cite{4}.
This fact had induced extensive theoretical activity mainly connected with
modeling the formation of quarkonium states from unbound heavy quark pairs 
produced in hard interaсtion. There exist two competing theoretical approaches
known in the literature under the names of color-singlet (CS) and color-octet 
(CO) models.
In general, a quark-antiquark pair is produced in a state $^{2S+1}L_J^{(a)}$
with spin $S$, orbital angular momentum $L$, total angular momentum $J$ and 
color $a$, which can be either identical to the respective quantum numbers of 
the resulting quarkonium (as accepted in the CS model) or different from those.
In the latter case, the heavy quark pair transforms into physical quarkonium 
state by means of soft (nonperturbative) gluon radiation, as considered in the
formalism of non-relativistic Quantum Chromodynamics (NRQCD)\cite{5,6}. The quarkonium
formation probability is then determined by the respective nonperturbative
matrix elements (NMEs), which are assumed to be universal (process-independent)
and not depending on the quarkonium momentum. Though not strictly calculable
within the theory, the NMEs are assumed to obey certain hierarchy in powers 
of the relative quark velocity $v$. To the leading order in $v$, an $S$-wave 
vector meson such as $\psi(2S)$ can be formed from a quark pair produced 
as color singlet $^3S_1^{(1)}$ or via one of intermediate color octet states 
$^1S_0^{(8)}$, $^3S_1^{(8)}$ or $^3P_J^{(8)}$ with $J = 0$, $1$ or $2$.

We know already that
the CS model with leading order (LO) hard scattering matrix elements fails
to describe the experimental data on $J/\psi$ and $\Upsilon$ production at the 
Tevatron and LHC. Including the next-to-leading order (NLO)\cite{7} 
and dominant tree-level next-to-next-to-leading order (NNLO$^*$)\cite{8}
corrections to the CS mechanism significantly improves the description of the 
collider data\cite{9}.
In the NRQCD formalism, a reasonably good agreement with the measured quarkonia 
cross sections can be achieved by adjusting the NME values which play the role 
of free fitting parameters\cite{10,11,12,13,14,15}. 
This was already demonstrated by comparing the calculations with the ATLAS\cite{16},
CMS\cite{17} and LHCb\cite{18} experimental data taken at $\sqrt s = 7$~TeV. However, the values 
of the extracted NMEs dramatically depend on the minimal quarkonium transverse 
momentum $p_T$ used in the fits\cite{19} and are incompatible with each other 
(both in size and even in sign!) when obtained from fitting the different sets 
of data.
Furthermore, none of the fits is able to accommodate the polarization 
measurements\cite{20,21}.
The fits involving low $p_T$ measurements lead to the conclusion that the 
production of $S$-wave quarkonia at high $p_T$ must be dominated by CO
contributions with transverse polarization (namely, by the $^3S_1^{(8)}$ channel). 
The latter contradicts to the unpolarized production seen by the CDF\cite{22,23}
Collaboration at the Tevatron\footnote{The CDF Collaboration has measured small 
longitudinal polarization in prompt $J/\psi$ production.} and CMS\cite{20,24} 
and LHCb\cite{21} Collaborations at the LHC. To obtain an unpolarized state 
it is necessary to either assume that quarkonium production is dominated by the 
scalar $^1S_0^{(8)}$ channel\cite{12} or restrict the NRQCD fits to very high 
$p_T$ region\cite{19}. This problem is known as "quarkonium polarization puzzle"
and still far from understanding. 

%A new solution to the spin alignment problem
A new solution to the polarization puzzle
has been guessed in the $k_T$-factorization approach of QCD\cite{25,26}, where studies 
of quarkonia production and polarization have their own 
long history (see, for example,\cite{27,28,29,30,31,32,33,34} and references therein).
The $k_T$-factorization approach is based on the Balitsky-Fadin-Kuraev-Lipatov 
(BFKL)\cite{35} or Ciafaloni-Catani-Fiorani-Marchesini (CCFM)\cite{36} gluon 
evolution equations and provides theoretical grounds for including the effects 
of initial gluon radiation and intrinsic gluon transverse 
momentum\footnote{A detailed description and discussion of the $k_T$-factorization
formalism can be found, for example, in reviews\cite{37}.}.
A combination of the usual CS scheme and the $k_T$-factorization formalism 
results in reasonably good description of the data on $J/\psi$, $\chi_c$ and 
$\Upsilon$ production at HERA\cite{27,29,30,33}, Tevatron\cite{28,32} and LHC\cite{34}. 
These results are also in good agreement with more complicated 
explicit NNLO$^*$ calculations\cite{7,8,9} performed in the collinear 
approximation of QCD. The longitudinal polarization of directly produced quarkonia 
in the $k_T$-factorization is an immediate consequence of initial gluon 
off-shellness\cite{27}. Adding the feed-down from $P$-wave states ($\chi_c$ and 
$\chi_b$) leads to essentially unpolarized prompt $J/\psi$ 
and $\Upsilon$ mesons\cite{32,34}.
In the $k_T$-factorization approach at LO, the $P$-wave states are produced in 
$2\to 1$ gluon-gluon fusion and are expected to dominate at high transverse 
momenta. However, the latest LHC data show that the feed-down contributions
$\chi_c\to J/\psi{+}\gamma$ and $\chi_b\to\Upsilon{+}\gamma$ do not constitute 
more than only $20 - 30$\% of the visible cross section at large $p_T$ values.

So, the production of $\chi_c$ and $\chi_b$ mesons requires a dedicated study
which will be the subject of our forthcoming papers, while in the present
analysis we concentrate on the direct mechanism and only restrict to $\psi(2S)$ 
mesons having no contamination from higher states.
Here we present a systematic analysis of ATLAS\cite{16}, CMS\cite{17} and LHCb\cite{18} data 
collected at $\sqrt s = 7$~TeV regarding the transverse momentum distributions
and polarization parameters $\lambda_\theta$, 
$\lambda_\phi$ and $\lambda_{\theta \phi}$ which describe the spin density matrix 
of the produced $\psi(2S)$ mesons.
%As it was noted above, polarization observables are essential
%in the understanding of quarkonium production in high energy hadron collisions. 

The outline of our paper is the following. In Section~2 we briefly recall the 
NRQCD formalism and the $k_T$-factorization approach. In Section~3 we perform 
a numerical fit to the latest CMS data and extract the color-octet NMEs using
three different sets of transverse momentum dependent (TMD) gluon distributions. 
Later in this section we check the compatibility of the extracted 
papameters with ATLAS and LHCb
data on the production and polarization of $\psi(2S)$ mesons. The comparison
is followed by a discussion. Our conclusions are collected in Section~4.

\section{Theoretical framework} \indent 

We start with briefly recalling the essential calculation steps. 
Our consideration is based on the following leading-order off-shell partonic 
subprocesses\cite{5,6}:
\begin{equation}
  g^*(k_1) + g^*(k_2) \to \psi^\prime \left[^3S_1^{(1)}\right](p) + g(k),
\end{equation}
\begin{equation}
  g^*(k_1) + g^*(k_2) \to \psi^\prime \left[^1S_0^{(8)}, \, ^3S_1^{(8)}, \, ^3P_J^{(8)}\right](p),
\end{equation}
\noindent 
where $J = 0, 1$ or $2$, and the four-momenta of all particles are indicated in 
parentheses. The subprocesses~(1) and~(2) represent the CS and CO contributions,
respectively. The corresponding production amplitudes can be obtained from the 
one for an unspecified $c\bar c$ state by applying the appropriate projection 
operators, which guarantee the proper quantum numbers of the $c\bar c$ state 
under consideration. These operators for the different 
spin and orbital angular momentum states read\cite{4}:
\begin{equation}
  \Pi\left[^1S_0\right] = \gamma_5 \left( \hat p_c + m_c\right)/m^{1/2},
\end{equation}
\begin{equation}
  \Pi\left[^3S_1\right] = \hat \epsilon(S_z) \left( \hat p_c + m_c\right)/m^{1/2},
\end{equation}
\begin{equation}
  \Pi\left[^3P_J\right] = \left( \hat p_{\bar c} - m_c\right) \hat \epsilon(S_z) \left( \hat p_c + m_c\right)/m^{3/2},
\end{equation}
\noindent 
where $m=2m_c$ is the mass of the considered $c\bar c$ state, and $p_c$ 
and $p_{\bar c}$ are the four-momenta of the charmed quark and anti-quark. 
%In accordance with the non-relativistic bound state formalism, the charmed
%quark mass $m_c$ is always set equal to $1/2$ of the quarkonium mass.
States with various projections of the spin momentum onto the $z$ axis are 
represented by the polarization four-vector $\epsilon_\mu(S_z)$.

The probability for the two quarks to form a meson depends on the bound state 
CS and fictituous CO wave functions $\Psi^{(a)}(q)$, where the relative 
four-momentum $q$ of the quarks in the bound state is treated as a small quantity
in the non-relativistic approximation. So, we represent the quark momenta as
\begin{equation}
  p_c = p/2 + q, \, p_{\bar c} = p/2 - q.
\end{equation}
\noindent 
Then, we multiply the hard subprocess amplitude ${\cal A}$ (depending on $q$) 
by the meson wave function $\Psi^{(a)}(q)$ and perform integration with respect
to $q$. The integration is done after expanding the amplitude ${\cal A}$ around $q = 0$:
\begin{equation}
  {\cal A}(q)\Psi^{(a)}(q) = {\cal A}|_{q = 0}\,\Psi^{(a)}(q) + q^\alpha (\partial {\cal A} / \partial q^\alpha)|_{q = 0}\,\Psi^{(a)}(q) + ...
\end{equation}
\noindent 
Since the expressions for ${\cal A}|_{q = 0}$ and 
$\partial {\cal A} / \partial q^\alpha|_{q = 0}$ are
no longer dependent on $q$, they may be factored outside the
integral sign. 
A term-by-term integration of this series employs the identities\cite{4}:
\begin{equation}
  \int {d^3 q\over (2 \pi)^3} \, \Psi^{(a)}(q) = {1\over \sqrt{4\pi}} {\cal R}^{(a)}(0), 
\end{equation}
\begin{equation}
  \int {d^3 q\over (2 \pi)^3} \, q^\alpha \Psi^{(a)}(q) = - i \epsilon^\alpha(L_z){\sqrt 3\over \sqrt{4\pi}} {\cal R}^{\prime \, (a)}(0),
\end{equation}

\noindent 
where ${\cal R}^{(a)}(x)$ are the radial wave functions in the coordinate 
representation.The first term in~(7) contributes only to $S$-waves, but vanishes 
for $P$-waves. On the contrary, the second term contributes only to $P$-waves, 
but vanishes for $S$-waves. States with various projections of the orbital angular
momentum onto the $z$ axis are represented by the polarization four-vector 
$\epsilon_\mu(L_z)$.
%The CS and fictituous CO wave functions of $S$-wave states 
%are directly related to the NMEs:
The NMEs of $S$-wave states are directly related to the 
CS and fictituous CO wave functions:
\begin{equation}
 \langle{\cal O}^{\psi}\left[{^{2S + 1}L_J^{(a)}}\right]\rangle = 2 N_c (2J + 1) |{\cal R}^{(a)}(0)|^2 /4\pi,
\end{equation}

\noindent 
where $N_c=3$ and $J=1$. A similar relation holds for ${\cal R}^{\prime\,(a)}$ 
if $P$-wave states are involved. The CS wave function at the origin of 
coordinate space is known from the measured $\psi(2S)$ leptonic decay width. 
The color-octet NMEs are extracted from experimental data and obey the relation
\begin{equation}
 \langle{\cal O}^{\psi}\left[{^{3}P_J^{(8)}}\right]\rangle = (2J + 1) \, \langle{\cal O}^{\psi}\left[{^{3}P_0^{(8)}}\right]\rangle,
\end{equation}

\noindent 
coming from heavy quark spin symmetry at LO in $v$. The polarization vectors 
$\epsilon_\mu(S_z)$ and $\epsilon_\mu(L_z)$ are defined as explicit four-vectors. 
In the frame where the $z$ axis is oriented along the quarkonium momentum vector 
$p_\mu = (E,0,0,|{\mathbf p}|)$, these polarization 
vectors read
\begin{equation}
  \epsilon_\mu(\pm 1) = (0, \pm 1, i, 0)/\sqrt 2, \, \epsilon_\mu(0) = (|{\mathbf p}|, 0, 0, E)/m.
\end{equation}

\noindent 
The states with definite $S_z$ and $L_z$ are translated into states 
with definite total momentum $J$ and its projection $J_z$ using the Clebsch-Gordan coefficients:
\begin{equation}
  \epsilon_{\mu \nu}(J, J_z) = \sum_{S_z, \, L_z} \langle 1, L_z; 1, S_z | J, J_z \rangle \, \epsilon_\mu(S_z) \, \epsilon_\nu(L_z).
\end{equation}

\noindent 
Further evaluation of partonic amplitudes is straightforward
and is done using the algebraic manipulation system \textsc{form}\cite{38}.
Our results for perturbative production amplitudes squared and summed
over polarization states agree with ones\cite{39}.
Here we only mention several technical points. First, according to the 
$k_T$-factorization prescription\cite{25,26}, the summation over the incoming 
off-shell gluon polarizations is done using the gluon spin density matrix in the 
form
$\overline { \epsilon^\mu \epsilon^{*\, \nu} } = {\mathbf k}_T^{\mu} {\mathbf k}_T^{\nu}/{\mathbf k}_T^2$,
where ${\mathbf k}_T$ is the gluon transverse momentum orthogonal to the beam axis.
In the collinear QCD limit, when $|{\mathbf k}_T| \to 0$, this expression converges to 
the ordinary $\overline { \epsilon^\mu \epsilon^{*\, \nu} } = - g^{\mu \nu}/2$ 
after averaging over the azimuthal angle.
Second, the $\psi(2S)$ spin density matrix is expressed in terms of the
momenta $l_1$ and $l_2$ of the decay leptons and is taken as
\begin{equation}
  \sum \epsilon^\mu \epsilon^{*\, \nu} = 3 \left( l_1^\mu l_2^\nu + l_1^\nu l_2^\mu - {m^2\over 2} g^{\mu \nu} \right)/m^2.
\end{equation}

\noindent 
This expression is equivalent to the standard one 
$\sum \epsilon^\mu \epsilon^{*\, \nu} = - g^{\mu \nu} + p^\mu p^\nu/m^2$ 
but is better suited for studying the polarization observables 
because it gives access to the kinematic variables
describing the orientation of the decay plane.
In all other respects the evaluation follows the standard QCD Feynman rules. 

An important point in the NRQCD formalism is connected with 
the emission of soft gluons taking place after the hard interaction is over.
It is usually assumed that the emitted soft gluons bring away the unwanted
color and change other quantum numbers of the produced CO system, 
but do not carry any energy, thus keeping the kinematics intact.
However, such an emission contradicts to the basic
QCD property: soft gluons can never be radiated as they are confined.
In order that the quantum numbers get changed, one needs to radiate
a real gluon with some energy $E \sim \Lambda_{\rm QCD}$, giving us the confidence 
that we do not enter into the confinement or perturbative domains. 
When considering the gluon radiation from $^3P_1^{(8)}$ and $^3P_2^{(8)}$
states, we rely upon the dominance of electric dipole $E1$ transitions,
which is supported by the E835 experimental data\cite{40}.
The corresponding invariant amplitudes can be written as follows\cite{41}:
\begin{equation}
  {\cal A}(^3P_1^{(8)} \to \psi^\prime + g) = g_1 \, e^{\mu \nu \alpha \beta} k_\mu^{(g)} \epsilon_\nu^{\rm (CO)} \epsilon_\alpha^{(\psi^\prime)} \epsilon_\beta^{(g)},
\end{equation}
\begin{equation}
  {\cal A}(^3P_2^{(8)} \to \psi^\prime + g) = g_2 \, p_\mu^{\rm (CO)} \epsilon_{\alpha \beta}^{\rm (CO)} \epsilon_\alpha^{(\psi^\prime)} \left[ k_\mu^{(g)} \epsilon_\beta^{(g)} - k_\beta^{(g)} \epsilon_\mu^{(g)} \right],
\end{equation}

\noindent 
where $p_\mu^{\rm (CO)}$, $k_\mu^{(g)}$, $\epsilon_\mu^{(\psi^\prime)}$, $\epsilon_\mu^{(g)}$,
$\epsilon_\mu^{\rm (CO)}$ and $\epsilon_{\mu \nu}^{\rm (CO)}$ are the four-momenta and 
polarization four-vectors (tensor) of corresponding particles and
$e^{\mu \nu \alpha \beta}$ is the fully antisymmetric Levi-Civita tensor.
The gluon radiation from other CO states is generated according to the phase space.

The cross section of $\psi(2S)$ production 
at high energies in the $k_T$-factorization approach
is calculated as a convolution of the off-shell 
partonic cross sections and the TMD gluon densities in a proton. 
The contribution from the CS production mechanism
can be presented in the following form:
\begin{equation}
  \displaystyle \sigma(p p \to \psi^\prime + X) = \int {1\over 16\pi (x_1 x_2 s)^2 } \, f_g(x_1,{\mathbf k}_{1T}^2,\mu^2) f_g(x_2,{\mathbf k}_{2T}^2,\mu^2) \, \times \atop
  \displaystyle  \times \, |\bar {\cal A}(g^* + g^* \to \psi^\prime + g)|^2 \, d{\mathbf p}_{T}^2 d{\mathbf k}_{1T}^2 d{\mathbf k}_{2T}^2 dy dy_g \, {d\phi_1 \over 2\pi} {d\phi_2 \over 2\pi},
\end{equation}

\noindent
where $f_g(x,{\mathbf k}_{T}^2,\mu^2)$ is the TMD
gluon density, ${\mathbf p}_T$ and $y$ are the transverse momentum and rapidity
of produced $\psi(2S)$ meson, $y_g$ is the rapidity of outgoing gluon 
and $\sqrt s$ is the $pp$ center-of-mass energy.
The initial off-shell gluons have a fraction $x_1$ and $x_2$ 
of the parent protons longitudinal 
momenta, non-zero transverse momenta ${\mathbf k}_{1T}$ and 
${\mathbf k}_{2T}$ (${\mathbf k}_{1T}^2 = - k_{1T}^2 \neq 0$, 
${\mathbf k}_{2T}^2 = - k_{2T}^2 \neq 0$) and azimuthal angles
$\phi_1$ and $\phi_2$.
For the CO production we have:
\begin{equation}
  \displaystyle \sigma(p p \to \psi^\prime + X) = \int {2 \pi\over x_1 x_2 s \, F} \, f_g(x_1,{\mathbf k}_{1T}^2,\mu^2) f_g(x_2,{\mathbf k}_{2T}^2,\mu^2) \, \times \atop
  \displaystyle  \times \, |\bar {\cal A}(g^* + g^* \to \psi^\prime)|^2 \, d{\mathbf k}_{1T}^2 d{\mathbf k}_{2T}^2 dy \, {d\phi_1 \over 2\pi} {d\phi_2 \over 2\pi}.
\end{equation}
 
 \noindent
According to the general definition\cite{42}, the off-shell gluon flux factor 
in~(18) is defined\footnote{The dependence of numerical predictions 
on the different forms of flux factor has been studied in\cite{31}.} 
as $F = 2 \lambda^{1/2}(\hat s,k_1^2,k_2^2)$, where
$\hat s = (k_1 + k_2)^2$. 
The squares of the corresponding off-shell partonic amplitudes, as 
being too lengthy, are not presented there but the full C++ code 
is available on request\footnote{lipatov@theory.sinp.msu.ru}.
The multidimensional integration have been performed by means of the Monte 
Carlo technique, using the routine \textsc{vegas}\cite{43}.

Numerically, we have tested several different sets of the TMD gluon densities.
Two of them (namely, JH and A0 sets) have been obtained\cite{44,45} 
from the CCFM equation where all input parameters have been fitted to 
describe the proton structure function $F_2(x, Q^2)$.
Besides the CCFM-evolved gluon densities, we applied the one 
obtained from the Kimber-Martin-Ryskin (KMR) prescription\cite{46}. 
The KMR approach is a formalism to construct
the TMD quark and gluon distributions from well-known conventional ones. 
For the input, we have used leading-order Martin-Stirling-Thorn-Watt (MSTW'2008) 
set\cite{47}.

\section{Numerical results} \indent

We now are in a position to present our numerical results. First we describe our
input and the kinematic conditions. Having the TMD gluon distributions chosen,
the cross sections~(17) and~(18) depend on the renormalization and factorization 
scales $\mu_R$ and $\mu_F$. We set 
$\mu_R^2 = m^2 + {\mathbf p}_{T}^2$ and
$\mu_F^2 = \hat s + {\mathbf Q}_T^2$, where ${\mathbf Q}_T$ is the 
transverse momentum of the initial off-shell gluon pair. 
The choice of $\mu_R$ is the standard one for studying the 
charmonia production whereas the special choice of $\mu_F$ is connected 
with the CCFM evolution\cite{44,45}.
Following\cite{48}, we set $\psi(2S)$ mass $m = 3.686$~GeV, 
branching fraction $B(\psi^\prime \to \mu^+ \mu^-) = 0.0077$ and use the LO formula 
for the coupling constant $\alpha_s(\mu^2)$ with $n_f = 4$ quark flavours
at $\Lambda_{\rm QCD} = 200$~MeV, such that $\alpha_s(M_Z^2) = 0.1232$.

In Table~1 we list our results for the NMEs fits obtained for three different
TMD gluon densities. We have fitted the transverse momentum distributions for 
prompt $\psi(2S)$ mesons measured recently by the CMS Collaboration at 
$\sqrt s = 7$~TeV\cite{17}. These measurements were done at moderate and high 
transverse momenta $10 < p_T < 100$~GeV, where the NRQCD formalism is believed 
to be most reliable. In contrast with\cite{11,15}, we performed the fitting 
procedure under requirement that the NME values be positive only. The color 
singlet NMEs were not fitted, but just taken from the known
$\psi(2S) \to \mu^+ \mu^-$ partial decay width\cite{48}. For comparison, 
we also present in Table~1 two sets of NMEs, obtained within the 
NLO NRQCD in\cite{11,15}. The main difference betwen\cite{11} 
and\cite{15} is in that these fits were based on differently selected sets of 
data points.

In the $k_T$-factorization approach, the fitted NME values strongly depend on 
the choice of TMD gluon density. We find that the $^1S_0^{(8)}$ contribution
is compatible with zero if the CCFM-evolved gluon distributions are used,
but is non-negligible in the case of KMR gluons.
%\footnote{This issue has to be traced to the different ${\mathbf k}_T^2$ 
%behaviour of these TMD gluon densities.}.
For the latter, the extracted value of 
$\langle {\cal O}^{\psi}\left[ ^1S_0^{(8)}\right] \rangle$
is very close to the one obtained in the NLO NRQCD analysis\cite{11}.
Both the NLO NRQCD fits\cite{11,15}
significantly (by one order of magnitude) 
exceeds the values of $\langle {\cal O}^{\psi}\left[ ^3S_1^{(8)}\right] \rangle$,
obtained with all of the TMD gluon densities. 
It is almost
consistent with estimates performed by other authors\cite{49, 50}.
%These NMEs are rather close to the
%estimates presented by other authors\cite{49, 50}.
In contrast with the results\cite{39}, our fit leads to non-zero 
$\langle {\cal O}^{\psi}\left[ ^3P_0^{(8)}\right] \rangle$ values.
Summing up, we can conclude that the NME values obtained by the different
authors on the basis of different data sets or by the same authors using
different gluon densities are widely spread, that spoils the belief in the
universality of the matrix elements.

\begin{table}
\begin{center}
\begin{tabular}{|c|c|c|c|c|}
\hline
  & & & & \\
    & $\langle {\cal O}^{\psi}\left[ ^3S_1^{(1)}\right] \rangle$/GeV$^3$ & $\langle {\cal O}^{\psi}\left[ ^1S_0^{(8)}\right] \rangle$/GeV$^3$ & $\langle {\cal O}^{\psi}\left[ ^3S_1^{(8)}\right] \rangle$/GeV$^3$ & $\langle {\cal O}^{\psi}\left[ ^3P_0^{(8)}\right] \rangle$/GeV$^5$ \\
  & & & & \\
\hline
  & & & & \\
  A0 & $7.04 \times 10^{-1}$ & 0.0 & $5.64 \times 10^{-4}$ & $3.71 \times 10^{-3}$ \\
  & & & & \\
  JH & $7.04 \times 10^{-1}$ & 0.0 & $3.19 \times 10^{-4}$ & $7.14 \times 10^{-3}$ \\
  & & & & \\
  KMR & $7.04 \times 10^{-1}$ & $8.14 \times 10^{-3}$ & $2.58 \times 10^{-4}$ & $1.19 \times 10^{-3}$ \\
  & & & & \\
  \cite{11} & $6.50 \times 10^{-1}$ & $7.01 \times 10^{-3}$ & $1.88 \times 10^{-3}$ & $-2.08 \times 10^{-3}$ \\
  & & & & \\
  \cite{15} & $5.29 \times 10^{-1}$ & $-1.20 \times 10^{-4}$ & $3.40 \times 10^{-3}$ & $9.45 \times 10^{-3}$ \\
  & & & & \\
\hline
\end{tabular}
\caption{The NMEs for $\psi(2S)$ meson derived from the fit of the CMS data\cite{17}. The NMEs
obtained in the NLO NRQCD fits\cite{11,15} are shown for comparison.}
\label{table1}
\end{center}
\end{table}

Now we turn to comparing our predictions with the  data collected by the ATLAS\cite{16},
CMS\cite{17} and LHCb\cite{18} Collaborations.
The ATLAS Collaboration has measured prompt $\psi(2S)$ transverse momentum 
distribution at $10 < p_T < 100$~GeV at central rapidities $|y| < 2$\cite{16}.
The CMS Collaboration probes the transverse momentum in the range $10 < p_T < 100$~GeV at
$|y| < 1.2$\cite{17}, and the LHCb Collaboration works in the kinematic 
range $p_T < 16$~GeV and $2 < y < 4.5$\cite{18}. 
In all cases the data were obtained at $\sqrt s = 7$~TeV.
The results of our calculations are shown in Figs.~1 --- 3.
With our sets of NMEs, we achieve reasonably good description of the data
with any of the considered TMD gluon density.
We observe dominance of the CO contributions in the whole $p_T$ range.
In particular, the $^3S_1^{(8)}$ contribution dominates at the large 
$p_T > 25$~GeV, whereas the $^1S_0^{(8)}$ channel is mostly important at 
low $p_T$  values.
Taken solely, the CS contributions (even incorporated with the 
$k_T$-factorization) are unable to describe the data.
They are important at relatively low $p_T$ only and are comparable there
with the $^3P_J^{(8)}$ contributions.
At moderate and high transverse momenta the CS contributions 
are below the data by about one order-of-magnitude.
The predictions obtained with the chosen TMD gluon densities are very close 
to each other at $p_T > 6$~GeV, while the difference becomes only sizable at 
low $p_T < 6$~GeV (see Fig.~3).
Therefore, similar to the collinear QCD factorization,
including the low $p_T$ data to the fit procedure 
can change the relative weight of different NMEs
in the $k_T$-factorization approach,
that is out of our present study.

%Similar to the collinear QCD factorization, including the low $p_T$ data can 
%change the relative weight of the different NMEs in the $k_T$-factorization 
%approach, that is out of our present study.

Now we turn to the the $\psi(2S)$ polarization issue, which is the most 
interesting part of our study.
In general, the spin density matrix of a vector particle decaying into a lepton
pair depends on three angular parameters $\lambda_{\theta}$, $\lambda_\phi$ and 
$\lambda_{\theta \phi}$ which can be measured experimentally. The double 
differential angular distribution of the decay leptons can be written as\cite{51}:
\begin{equation}
  {d\sigma \over d\cos \theta^* d\phi^*} \sim 1 + \lambda_\theta \cos^2 \theta^* + 
    \lambda_\phi \sin^2 \theta^* \cos 2 \phi^* + \lambda_{\theta \phi} \sin 2 \theta^* \cos \phi^*,
\end{equation}

\noindent
where $\theta^*$ and $\phi^*$ are the polar and azimuthal angles of the decay 
lepton measured in the charmonium rest frame. The case of $(\lambda_{\theta}, 
\lambda_\phi, \lambda_{\theta \phi}) = (0,0,0)$ corresponds to unpolarized state, 
while 
$(\lambda_{\theta}, \lambda_\phi, \lambda_{\theta \phi}) = (1,0,0)$ and
$(\lambda_{\theta}, \lambda_\phi, \lambda_{\theta \phi}) = (-1,0,0)$ refer 
to fully transverse and longitudinal polarizations. 
The CMS\cite{20} and LHCb\cite{21} Collaborations have measured all these 
parameters as functions of $\psi(2S)$ transverse momentum in two complementary 
frames: the Collins-Soper and helicity ones. 
In addition, the CMS Collaboration provided measurements in the perpendicular 
helicity frame.
In the Collins-Soper frame the polarization axis $z$ bisects the two beam 
directions whereas the polarization axis in the helicity frame
coincides with the direction of $\psi(2S)$ momentum in the laboratory frame.
In the perpendicular helicity frame the $z$ axis is orthogonal to that
in the Collins-Soper frame and lies in the plane spanned by the two beam
momenta. Additionally, the frame-independent polarization parameter\cite{51,52}
$\lambda^* = (\lambda_\theta + 3 \lambda_\phi)/(1 - \lambda_\phi)$
was investigated. Below we estimate the 
polarization parameters $\lambda_\theta$, $\lambda_\phi$, $\lambda_{\theta \phi}$
and $\lambda^*$ for the CMS and LHCb conditions.
Our calculation generally follows the experimental procedure. We collect 
the simulated events in the kinematical region defined by the CMS and LHCb 
experiments, generate the decay lepton angular distributions according
to the production and decay matrix elements, and then apply a three-parametric
fit based on~(19).

In Figs.~4 --- 8 we confront our predictions for polarization 
parameters $\lambda_\theta$, $\lambda_\phi$, 
$\lambda_{\theta \phi}$ and $\lambda^*$ with the latest 
CMS\cite{20} and LHCb\cite{21} data.
We find slight transverse polarization ($\lambda_\theta \sim 0.2$)
%of $\psi(2S)$ mesons
in the Collins-Soper frame and slight longitudinal 
polarization ($\lambda_\theta \sim - 0.2$) in the helicity frame
at low transverse momenta covered by the LHCb experiment.
These results are practically independent on the $\psi(2S)$ rapidity.
At higher $p_T$ the polarizations of $\psi(2S)$ mesons, calculated
in the Collins-Soper frame, go from slight transverse ($\lambda_\theta \sim 0.15$)
to almost zero values ($\lambda_\theta \sim 0.05$) as the transverse 
momentum increases from $p_T \sim 10$~GeV to $50$~GeV.
In the helicity and perpendicular helicity frames $\psi(2S)$ polarization
changes from longitudinal ($\lambda_\theta \sim - 0.2$)
to slight longitudinal ($\lambda_\theta \sim - 0.1$).
Here we arrive at the key point of our paper.
Figs.~4 --- 8 show that $\psi(2S)$ production,
calculated in the $k_T$-factorization approach,
tends to be unpolarized at high $p_T$, in agreement with the CMS data\cite{20}.
Indeed, as a strict consequence of the initial gluon off-shellness, a large 
fraction of $\psi(2S)$ mesons with zero helicity is produced in the partonic 
subprocesses, including both CS and CO contributions. 
Moreover, the fraction of such events increases when $p_T$ grows up (see, 
for example,\cite{27,28,29,30,31,32,33,34}).
The only exception refers to $^1S_0^{(8)}$ channel,
which is produced unpolarized due to its spinless nature.
It is a remarkable property of the $k_T$-factorization scheme that the gluon
fragmentation to $^3S_1^{(8)}$ states produces nearly unpolarized $\psi(2S)$
mesons (in contrast with conventional collinear NRQCD where the mesons carry
strong transverse polarization).
Thus, we can conclude that the problem of $\psi(2S)$ spin alignment can be 
solved if the initial gluon off-shellness is taken into account.

A comparison of our predictions with the LHC data\cite{20,21} shows
that the latter seem to support the trend observed in the $k_T$-factorization 
formalism.
However, while our predictions for $\lambda_\phi$ and 
$\lambda_{\theta \phi}$ parameters agree with the data,
the description of $\lambda_{\theta}$ and $\lambda^*$ 
is still rather qualitative than quantitative,
due to the huge experimental uncertainties.

Finally, we would like note that 
there are significant theoretical uncertainties 
connected with the choice of the renormalization and/or factorization
scales, the inclusion of NLO subprocesses and exact definition of NMEs.
The detailed study of these uncertainties is out of our present paper.

\section{Conclusions} \indent 

We have considered prompt $\psi(2S)$ production and polarization in $pp$ 
collisions at the LHC energy $\sqrt s = 7$~TeV in the framework 
of $k_T$-factorization approach. We have used the LO non-relativistic QCD
formalism including both color-singlet and color-octet contributions.
Using the TMD gluon densities in a proton derived 
from the CCFM equation and from the Kimber-Martin-Ryskin prescription,
we extracted the color-octet NMEs 
$\langle {\cal O}^{\psi}\left[ ^1S_0^{(8)}\right] \rangle$,
$\langle {\cal O}^{\psi}\left[ ^3S_1^{(8)}\right] \rangle$ and
$\langle {\cal O}^{\psi}\left[ ^3P_0^{(8)}\right] \rangle$ for $\psi(2S)$
mesons from fits to transverse momentum distributions provided by the latest CMS 
measurements.
Using the fitted NMEs, we have successfully described the data presented by 
the ATLAS, CMS and LHCb Collaborations. 
We estimated the polarization parameters $\lambda_\theta$, 
$\lambda_\phi$ and $\lambda_{\theta \phi}$ 
and demonstrated that taking into account the off-shellness
of the initial gluons in the color-octet contributions 
leads to unpolarized $\psi(2S)$ production at high
transverse momenta, that is in qualitative agreement with the LHC data.

\section{Acknowledgements} \indent 

The authors are grateful to H.~Jung
for very useful discussions and remarks.
This research was supported by the FASI of Russian Federation
(grant NS-3042.2014.2).
We are also grateful to DESY Directorate for the
support in the framework of Moscow---DESY project on Monte-Carlo implementation for
HERA---LHC.

\newpage

\begin{figure}
\begin{center}
\epsfig{figure=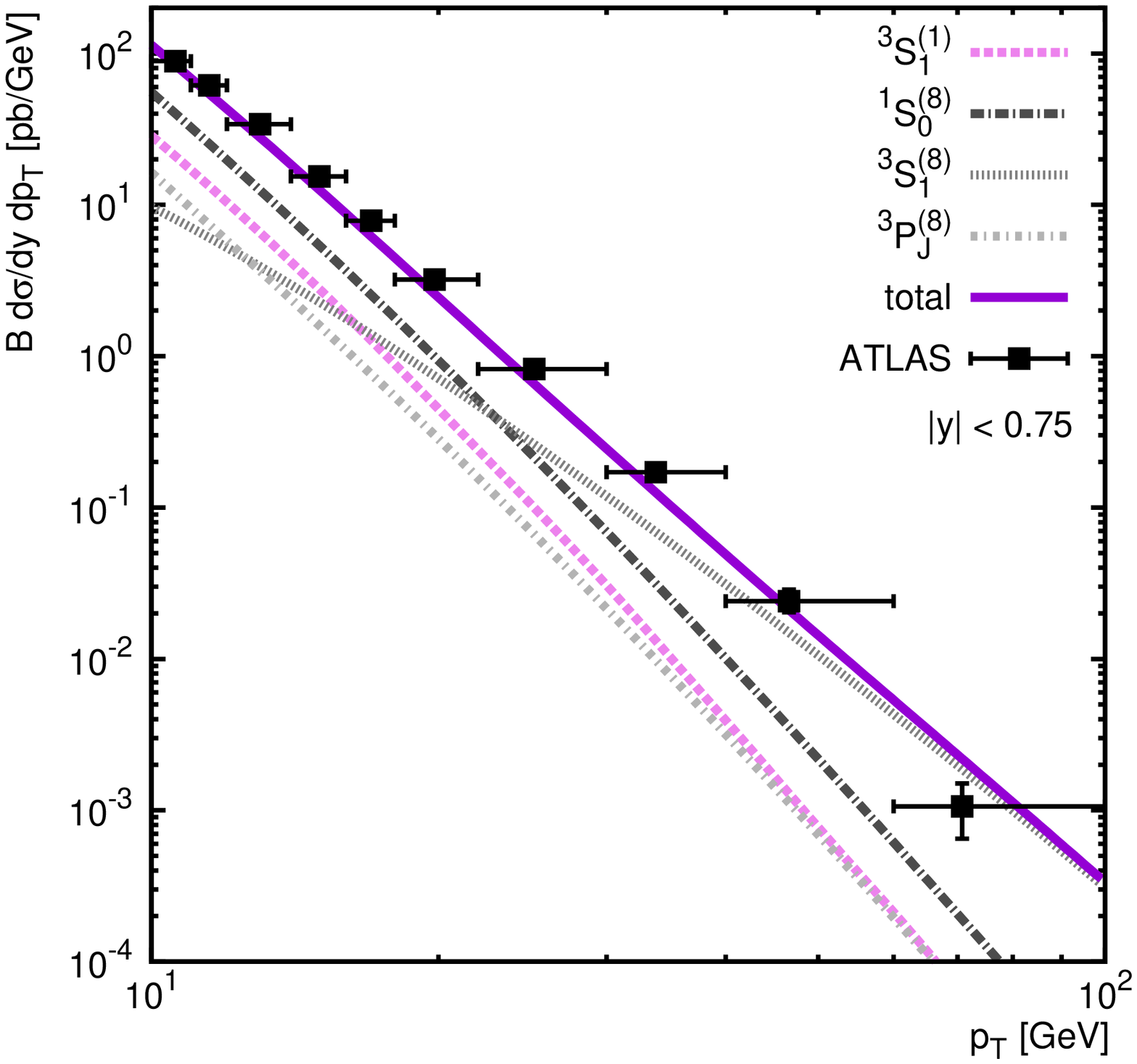, width = 6cm, angle = 270} 
\vspace{0.7cm} \hspace{-1cm}
\epsfig{figure=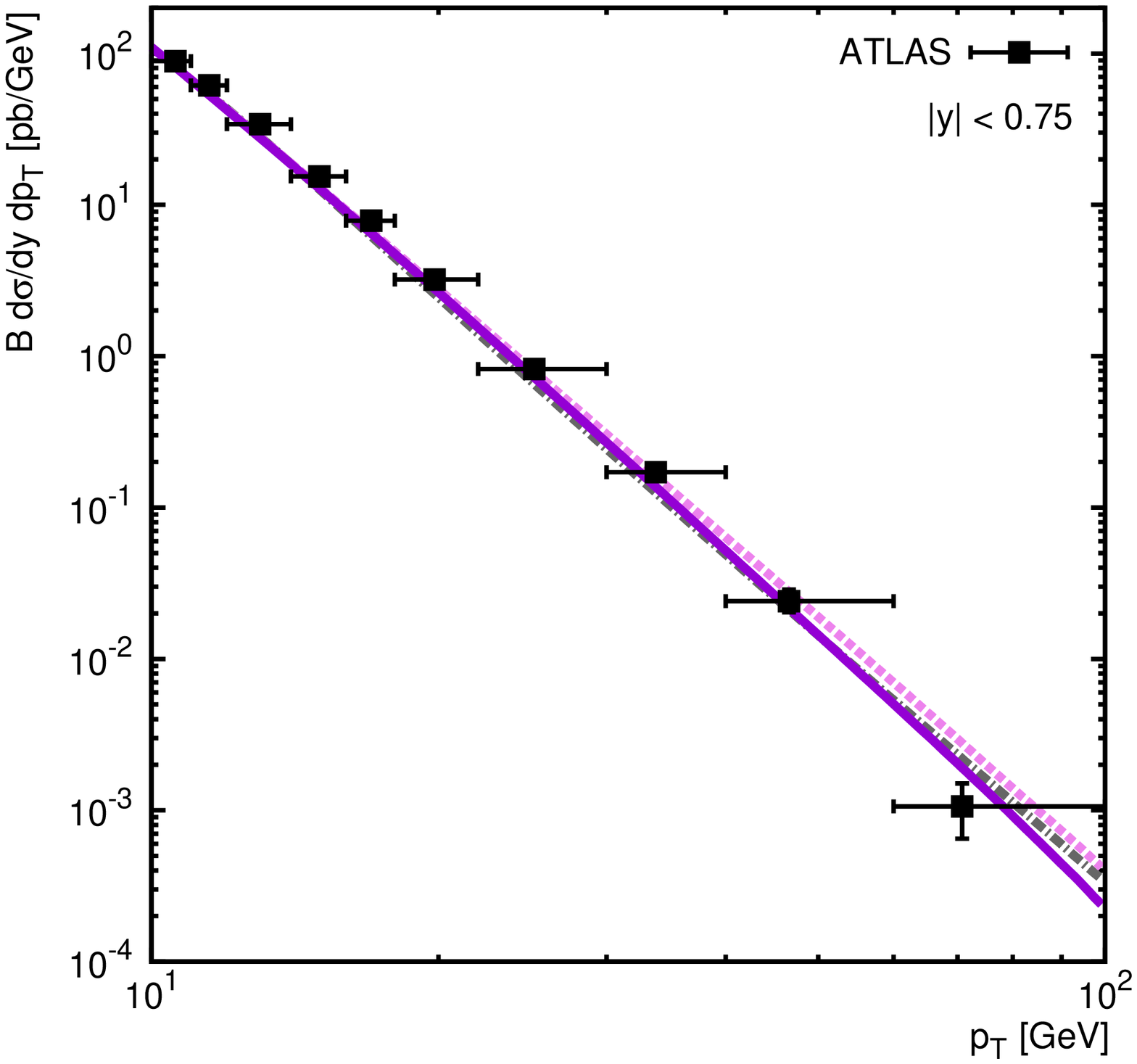, width = 6cm, angle = 270} 
\vspace{0.7cm}
\epsfig{figure=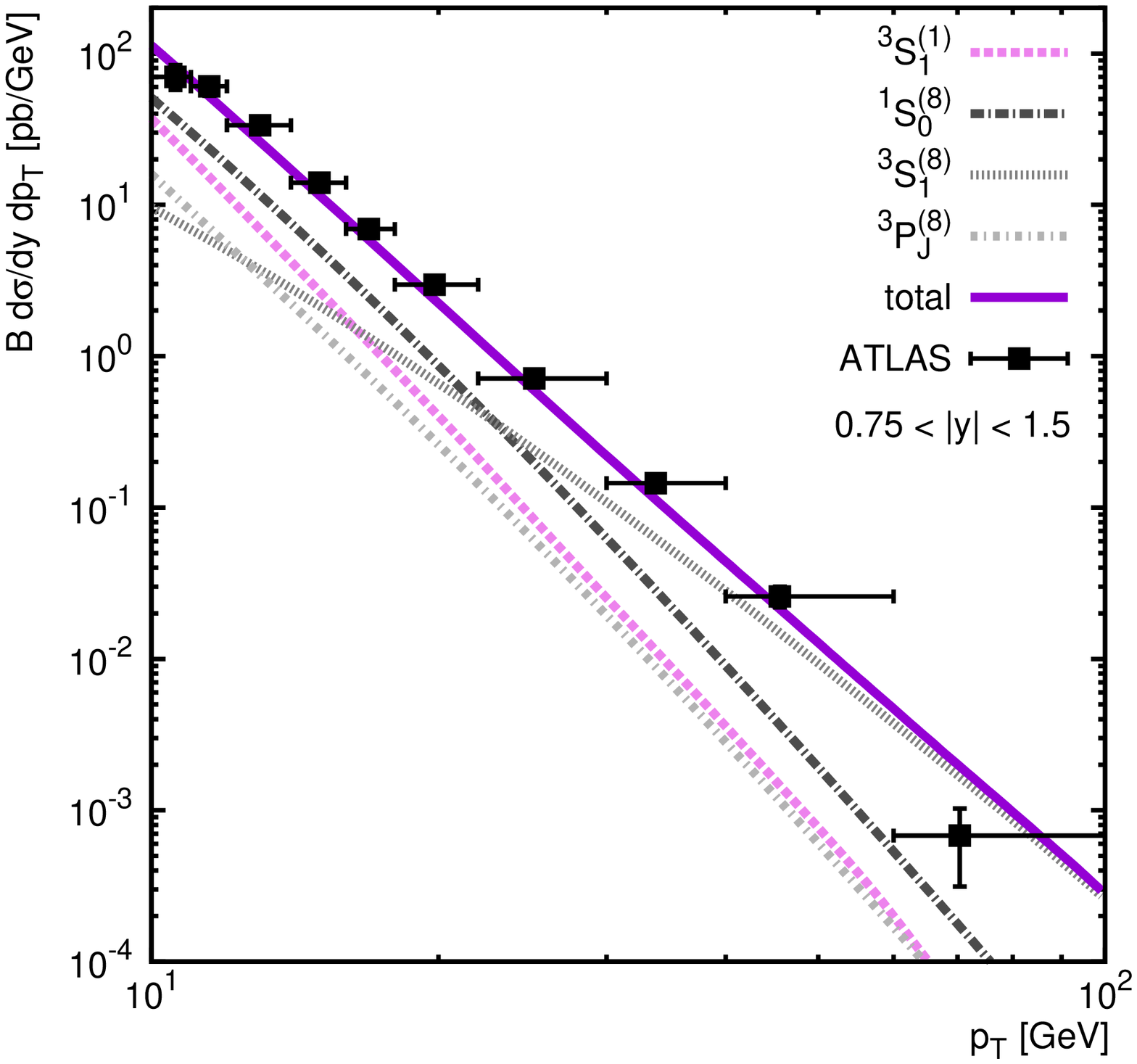, width = 6cm, angle = 270}
\hspace{-1cm}
\epsfig{figure=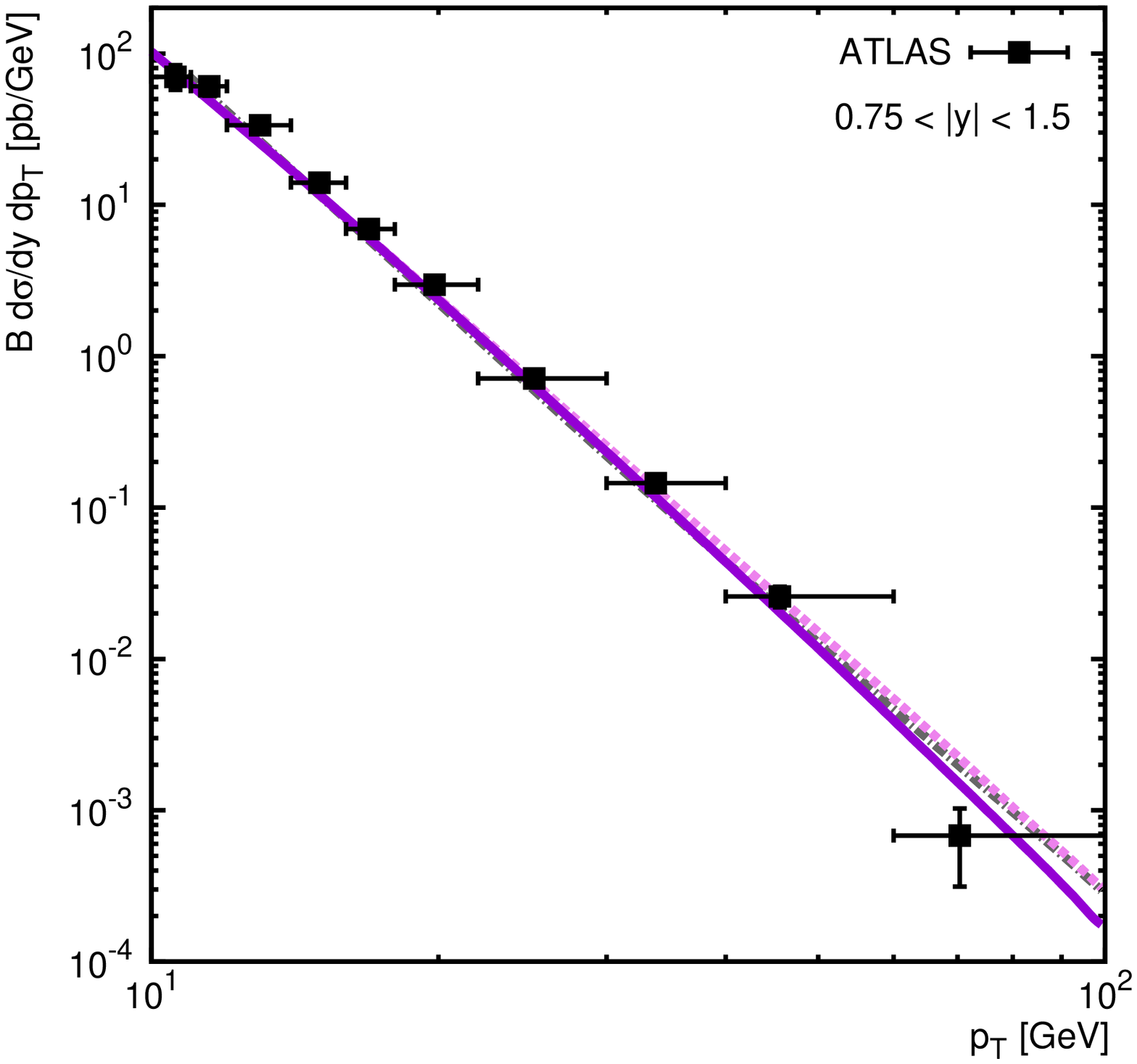, width = 6cm, angle = 270}
\epsfig{figure=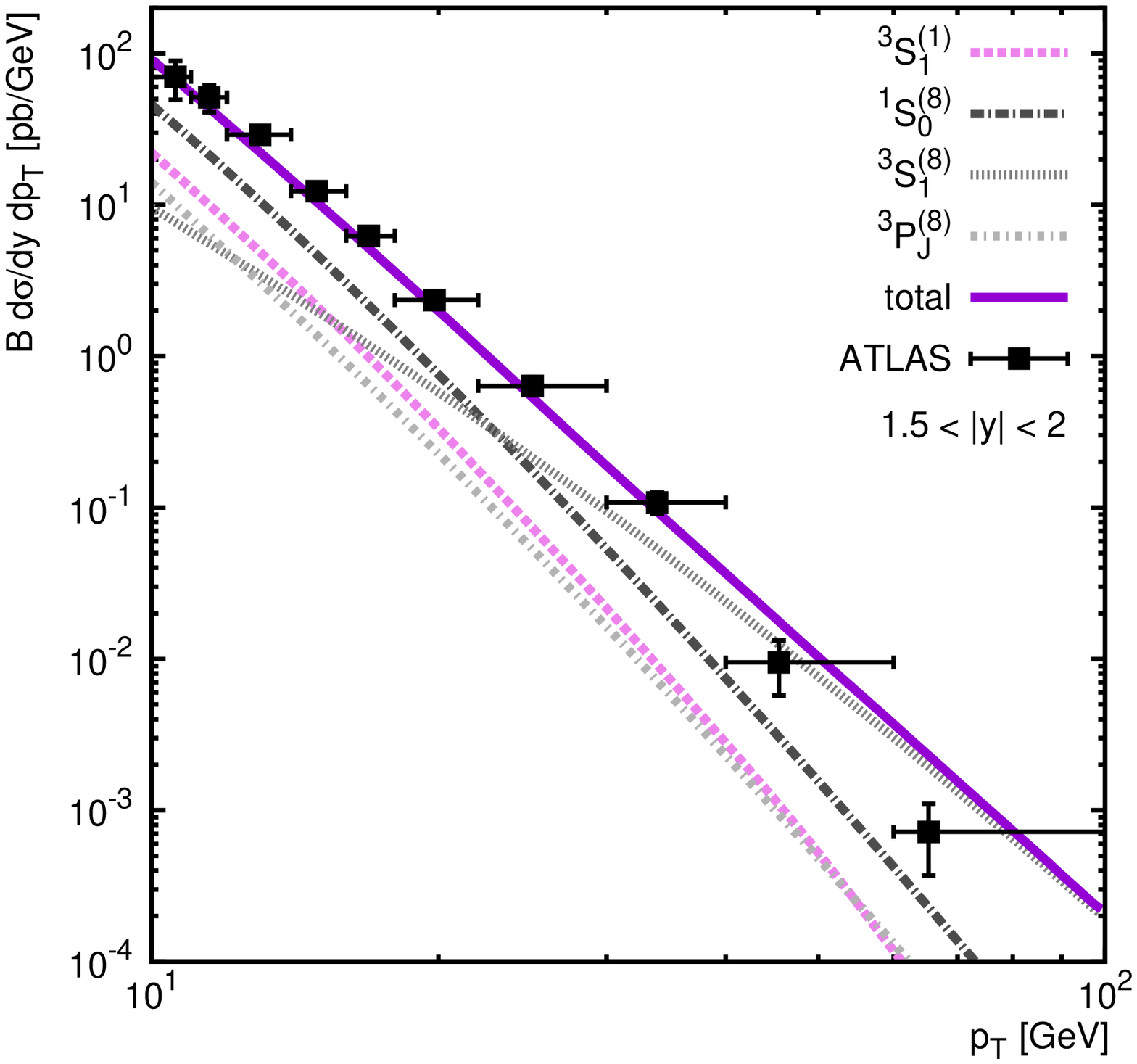, width = 6cm, angle = 270}
\hspace{-1cm}
\epsfig{figure=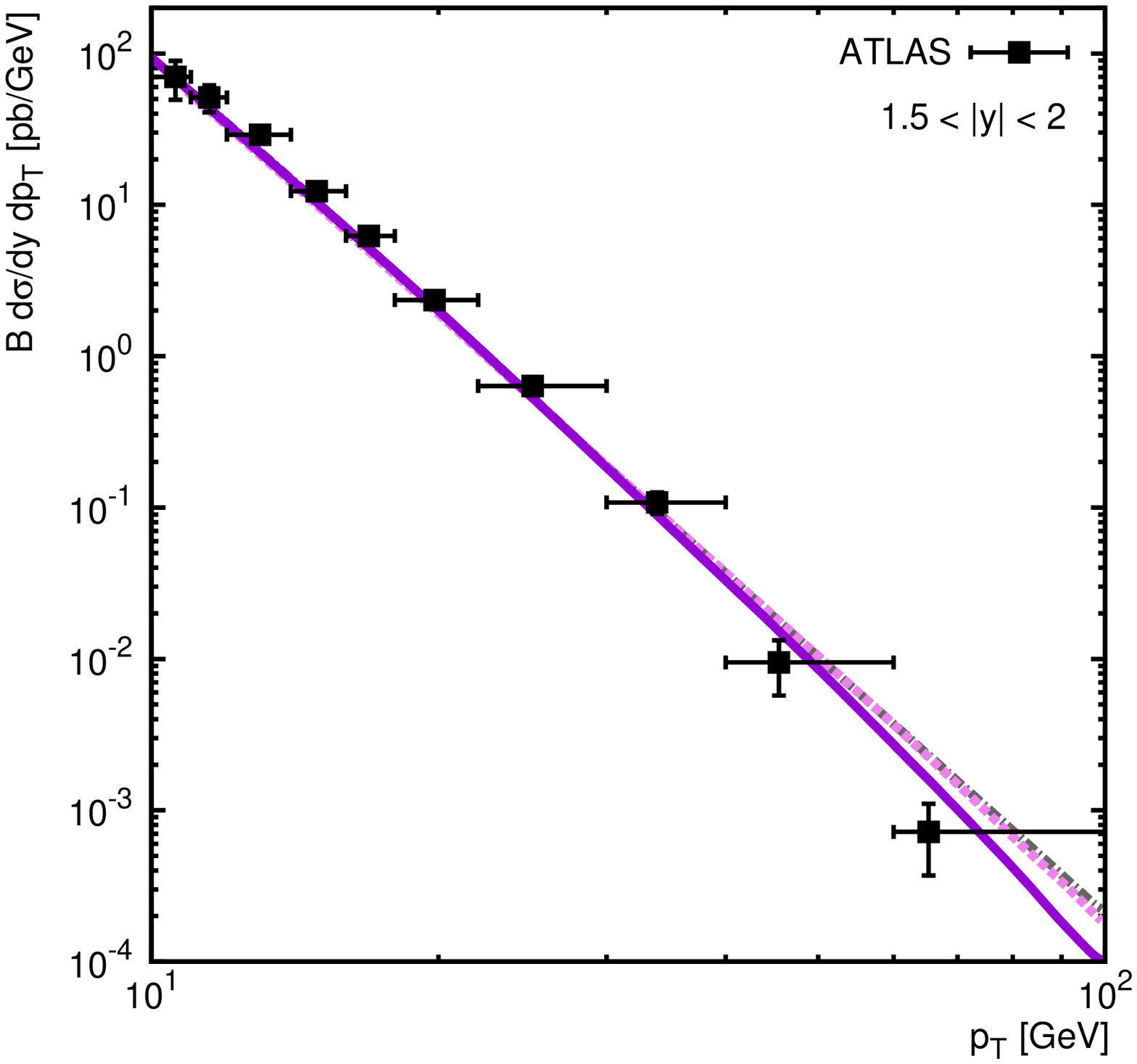, width = 6cm, angle = 270}
\caption{The double differential cross sections of prompt $\psi(2S)$ meson production
at the LHC.
Left panel: the dashed, dash-dotted, dotted and short dash-dotted
curves correspond to the color-singlet $^3S_1^{(1)}$ and color-octet 
$^1S_0^{(8)}$, $^3S_1^{(8)}$, $^3P_J^{(8)}$ contributions calculated with the KMR gluon density.
The solid curve represent the sum of CS and CO terms.
Right panel: the solid, dashed and dash-dotted curves correspond to the 
predictions obtained with the A0, JH and KMR gluon distributions, 
respectively. The experimental data are from ATLAS\cite{16}.}
\label{fig1}
\end{center}
\end{figure}

\begin{figure}
\begin{center}
\epsfig{figure=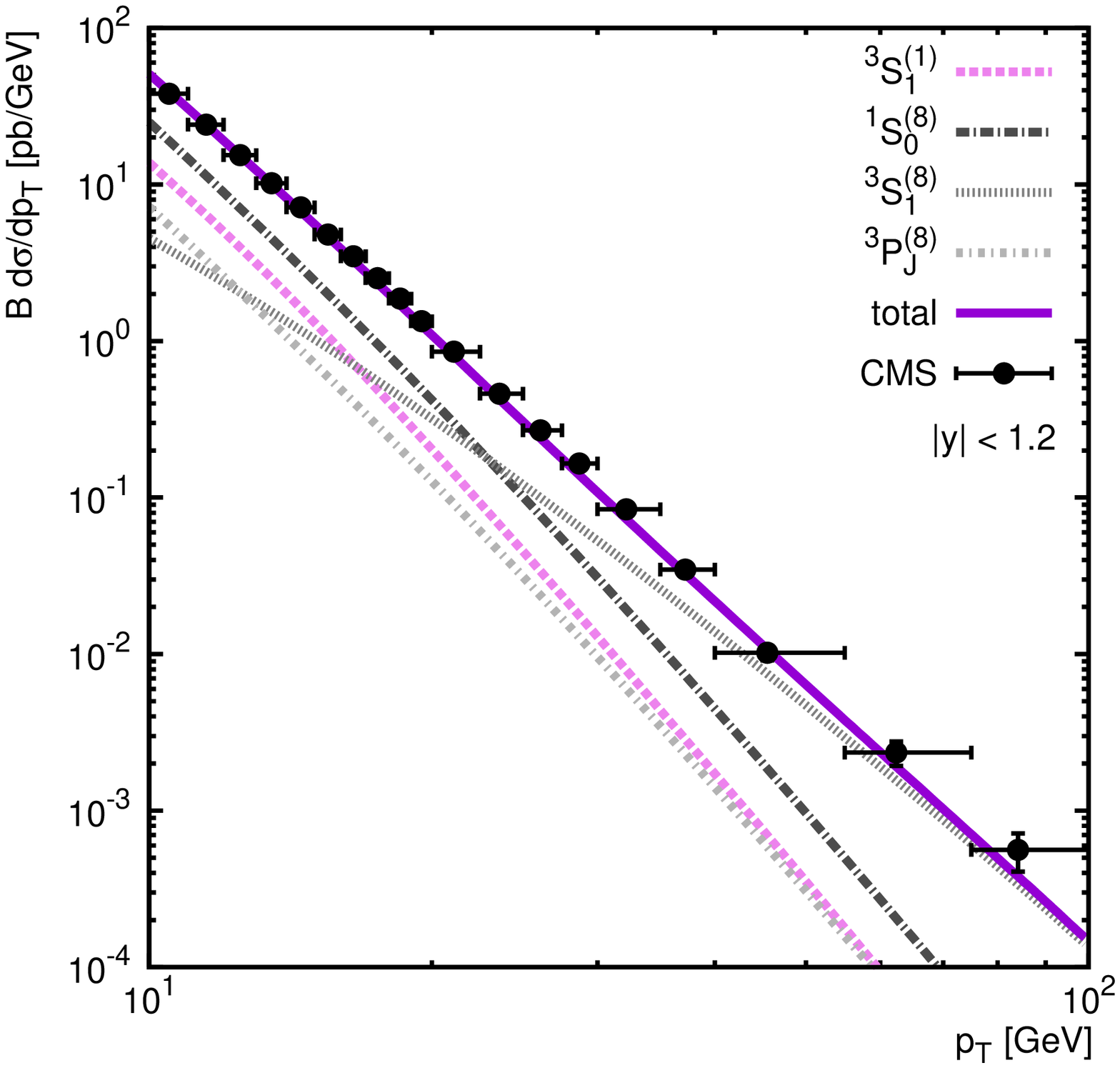, width = 6cm, angle = 270}
\hspace{-1cm}
\epsfig{figure=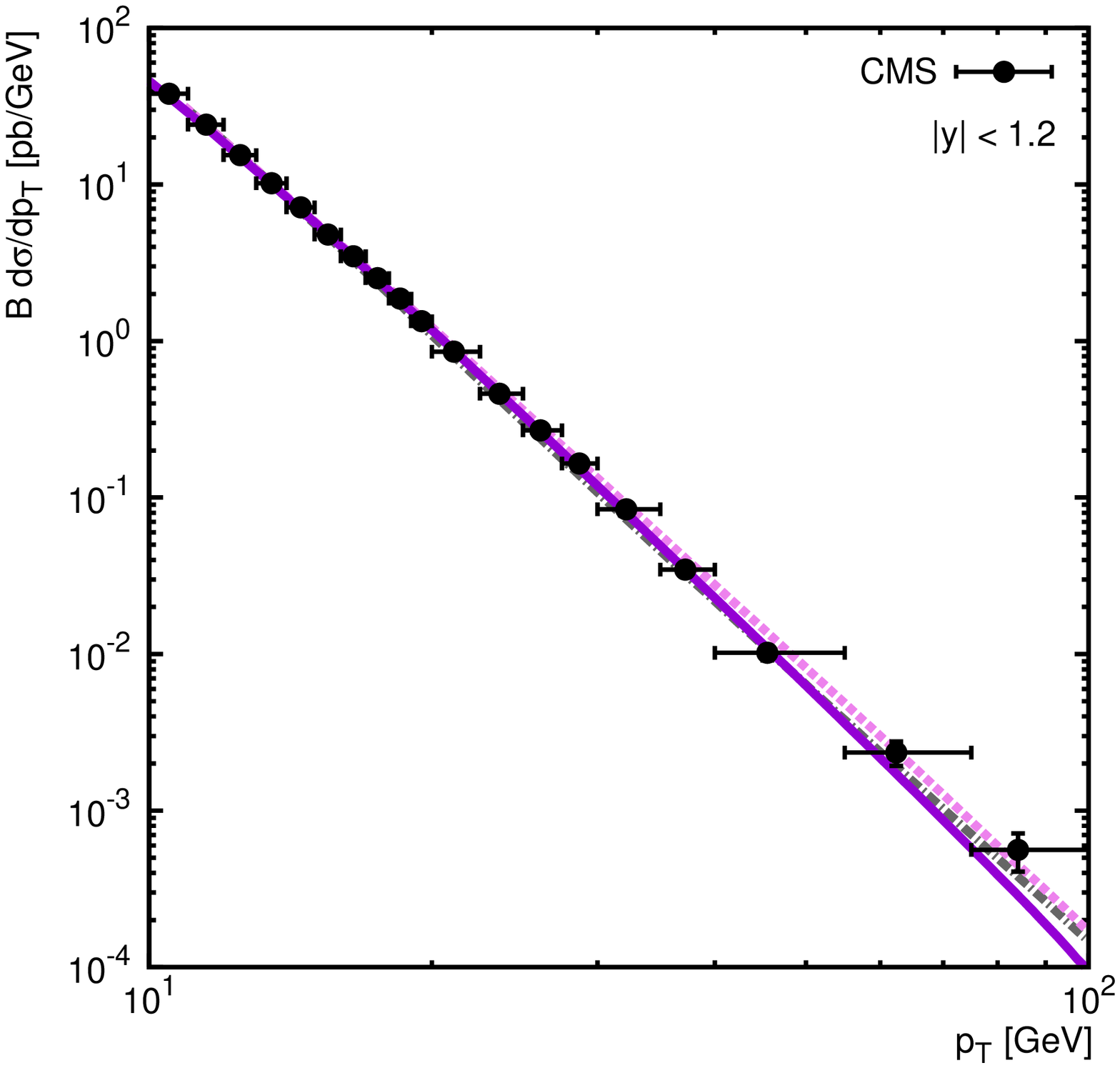, width = 6cm, angle = 270}
\caption{The transverse momentum distribution of prompt $\psi(2S)$ meson production
in $pp$ collisions at the LHC. 
Notation of all curves is the same as in Fig.~1. The experimental data are from CMS\cite{17}.}
\label{fig2}
\end{center}
\end{figure}

\begin{figure}
\begin{center}
\epsfig{figure=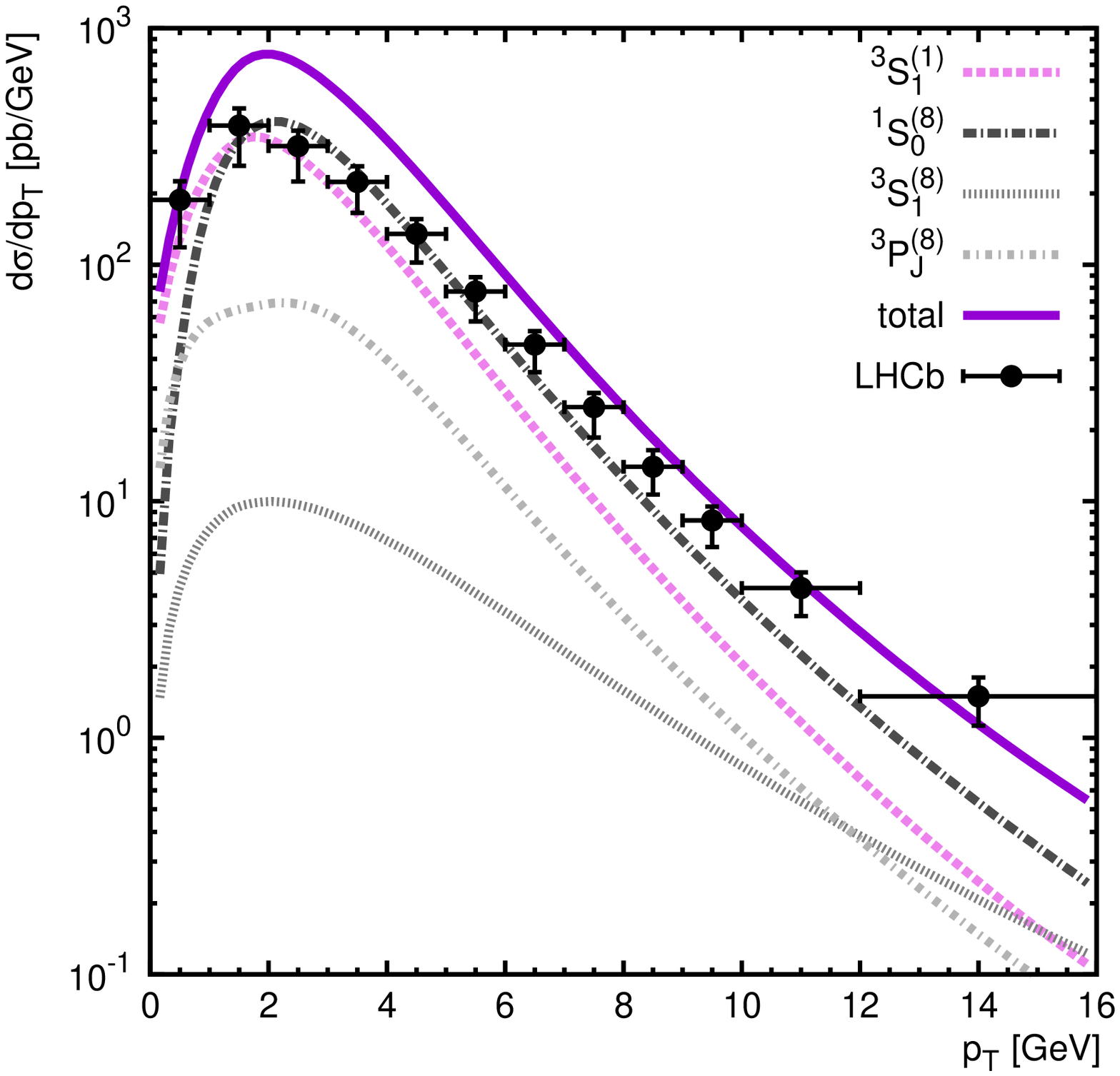, width = 6cm, angle = 270}
\hspace{-1cm}
\epsfig{figure=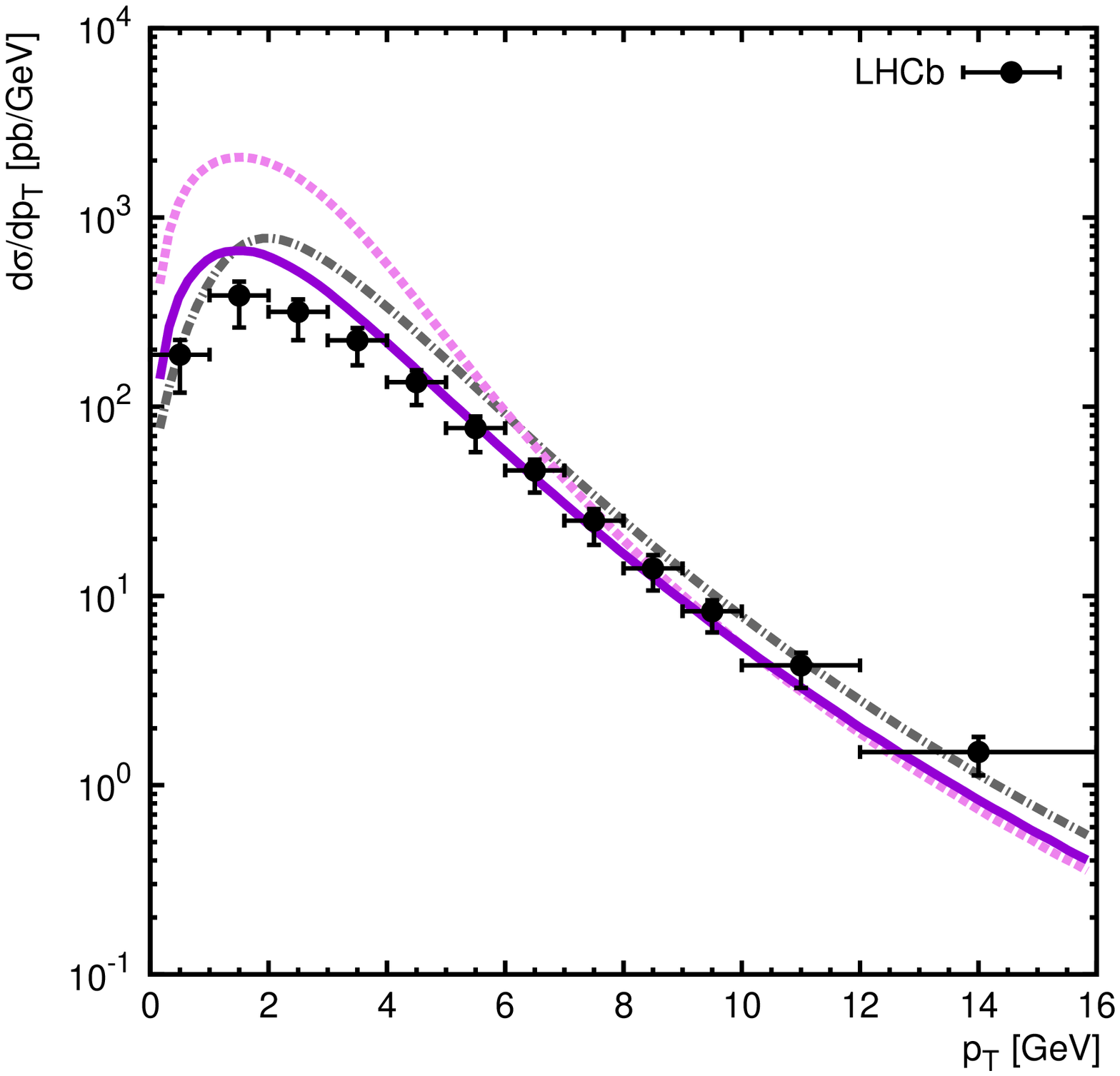, width = 6cm, angle = 270}
\caption{The transverse momentum distribution of prompt $\psi(2S)$ meson production
in $pp$ collisions at the LHC. 
Notation of all curves is the same as in Fig.~1. The experimental data are from LHCb\cite{18}.}
\label{fig3}
\end{center}
\end{figure}

\begin{figure}
\begin{center}
\epsfig{figure=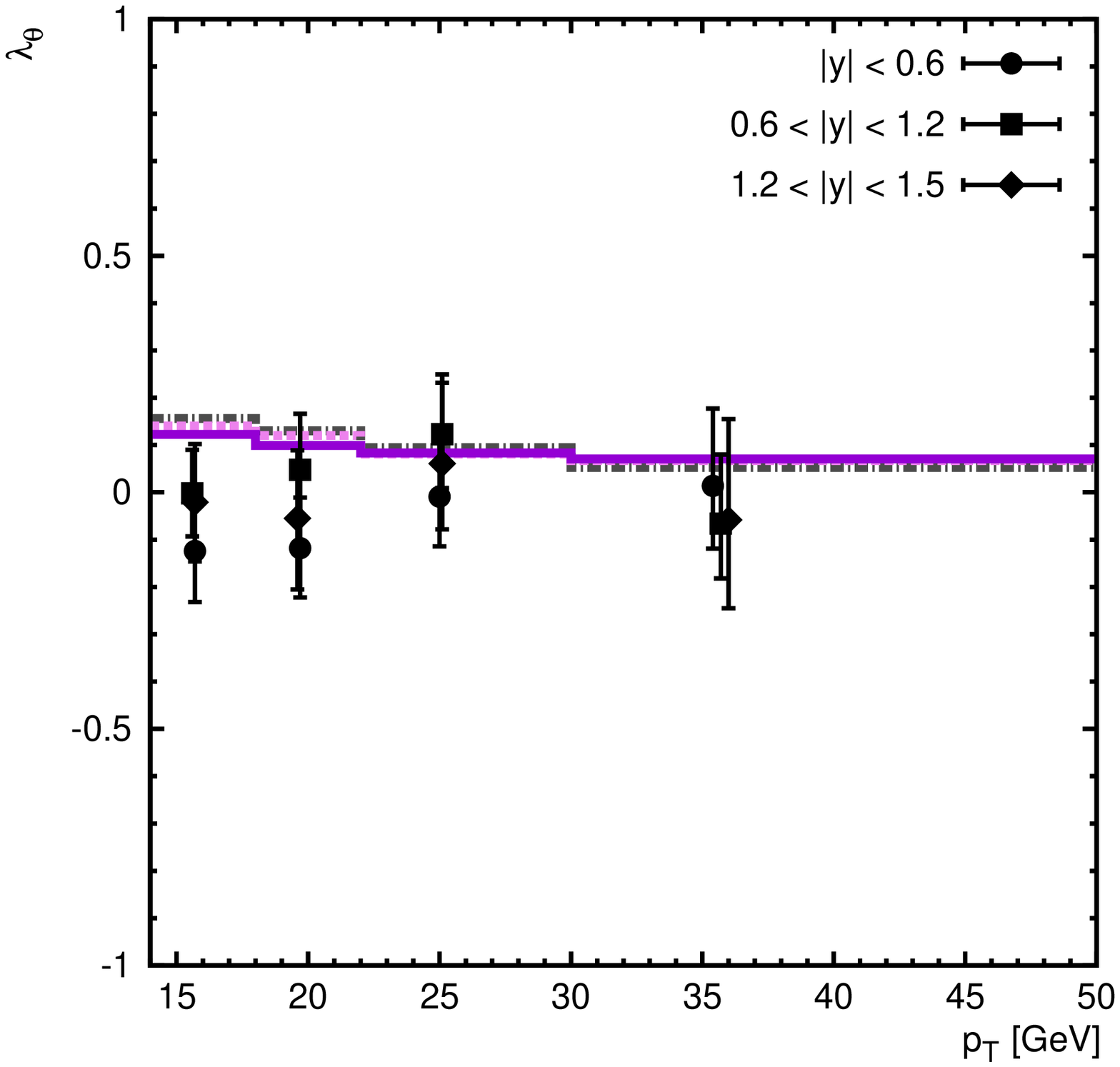, width = 6cm, angle = 270}
\vspace{0.7cm} \hspace{-1cm}
\epsfig{figure=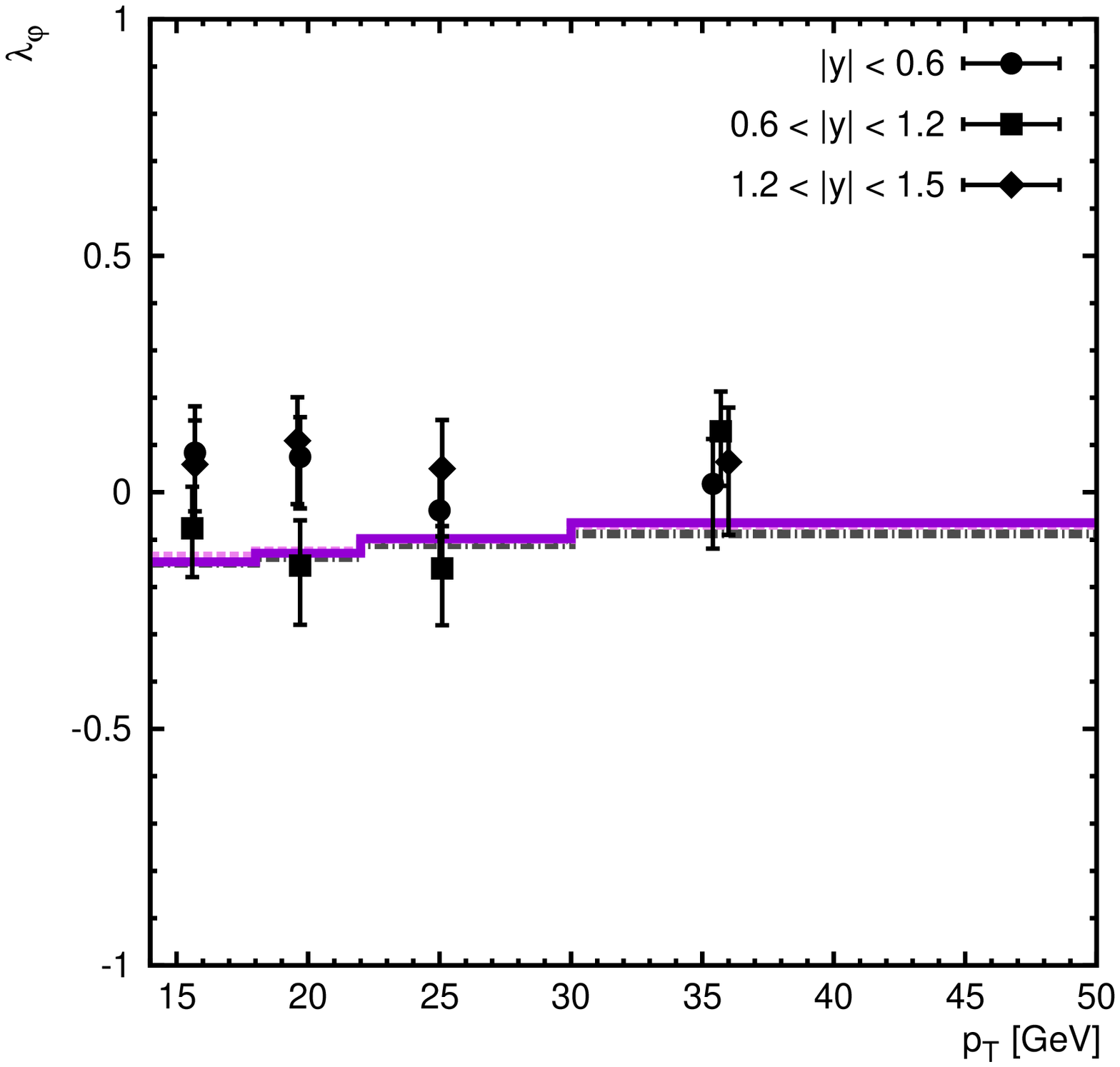, width = 6cm, angle = 270}
\epsfig{figure=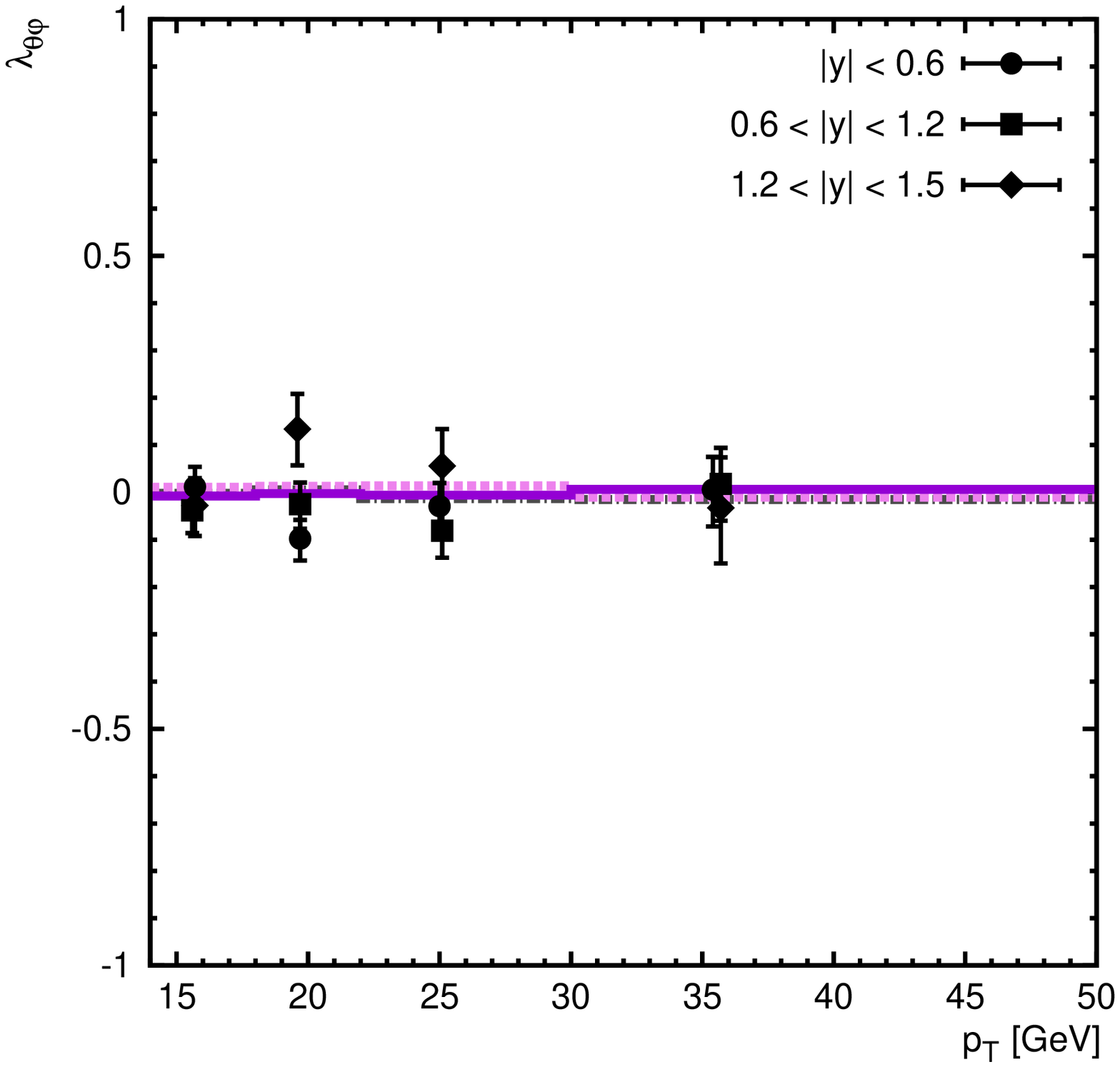, width = 6cm, angle = 270}
\hspace{-1cm} 
\epsfig{figure=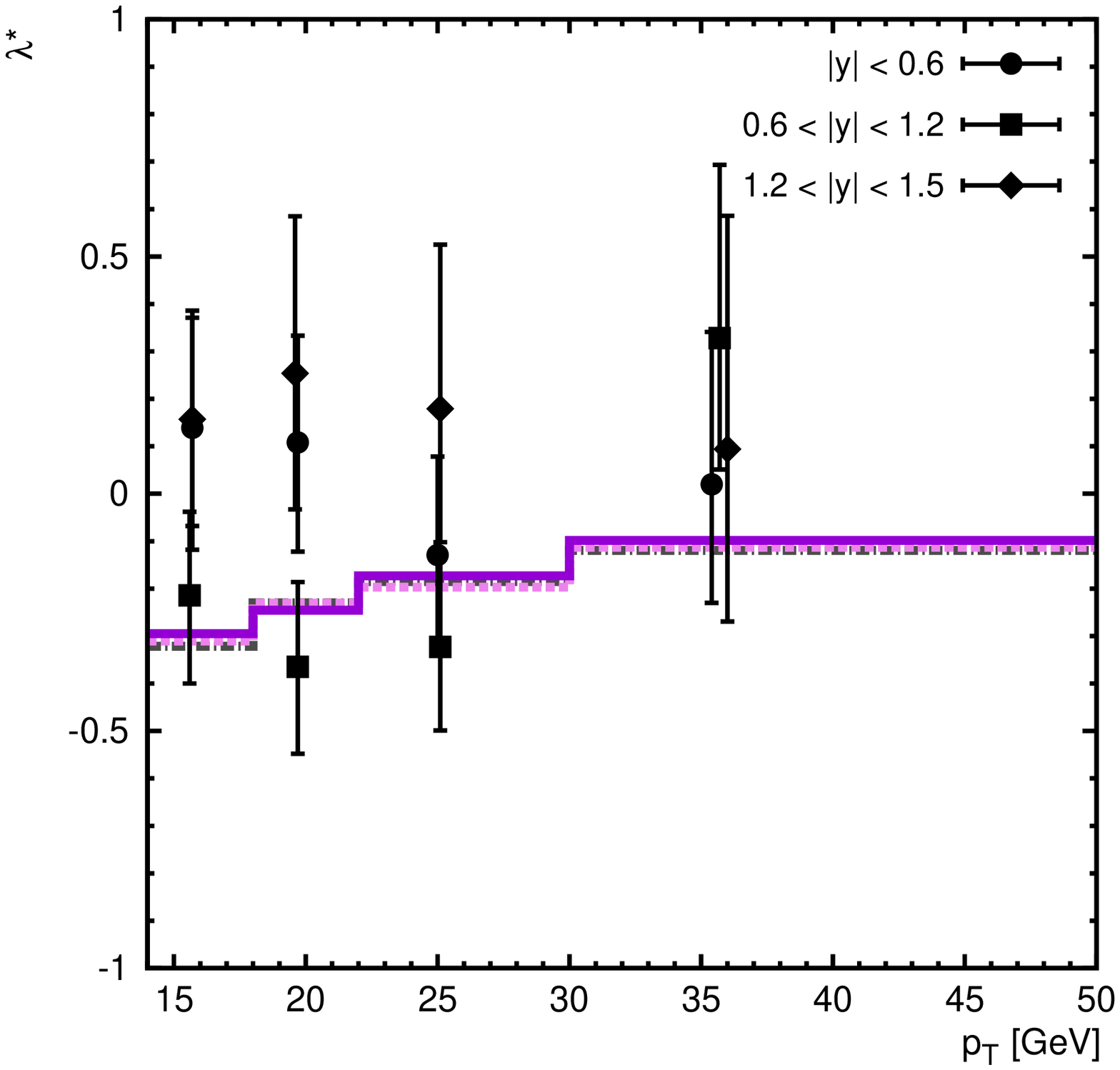, width = 6cm, angle = 270}
\caption{Polarization parameters 
$\lambda_\theta$, $\lambda_\phi$, $\lambda_{\theta \phi}$
and $\lambda^*$ of prompt $\psi(2S)$ mesons 
calculated as a function of $\psi(2S)$ transverse momentum in the Collins-Soper frame.
The solid, dashed and dash-dotted histograms correspond
to the predictions obtained at $|y| < 0.6$, $0.6 < |y| < 1.2$
and $1.2 < |y| < 1.5$. The KMR gluon distribution is used.
The experimental data are from CMS\cite{20}.}
\label{fig4}
\end{center}
\end{figure}

\begin{figure}
\begin{center}
\epsfig{figure=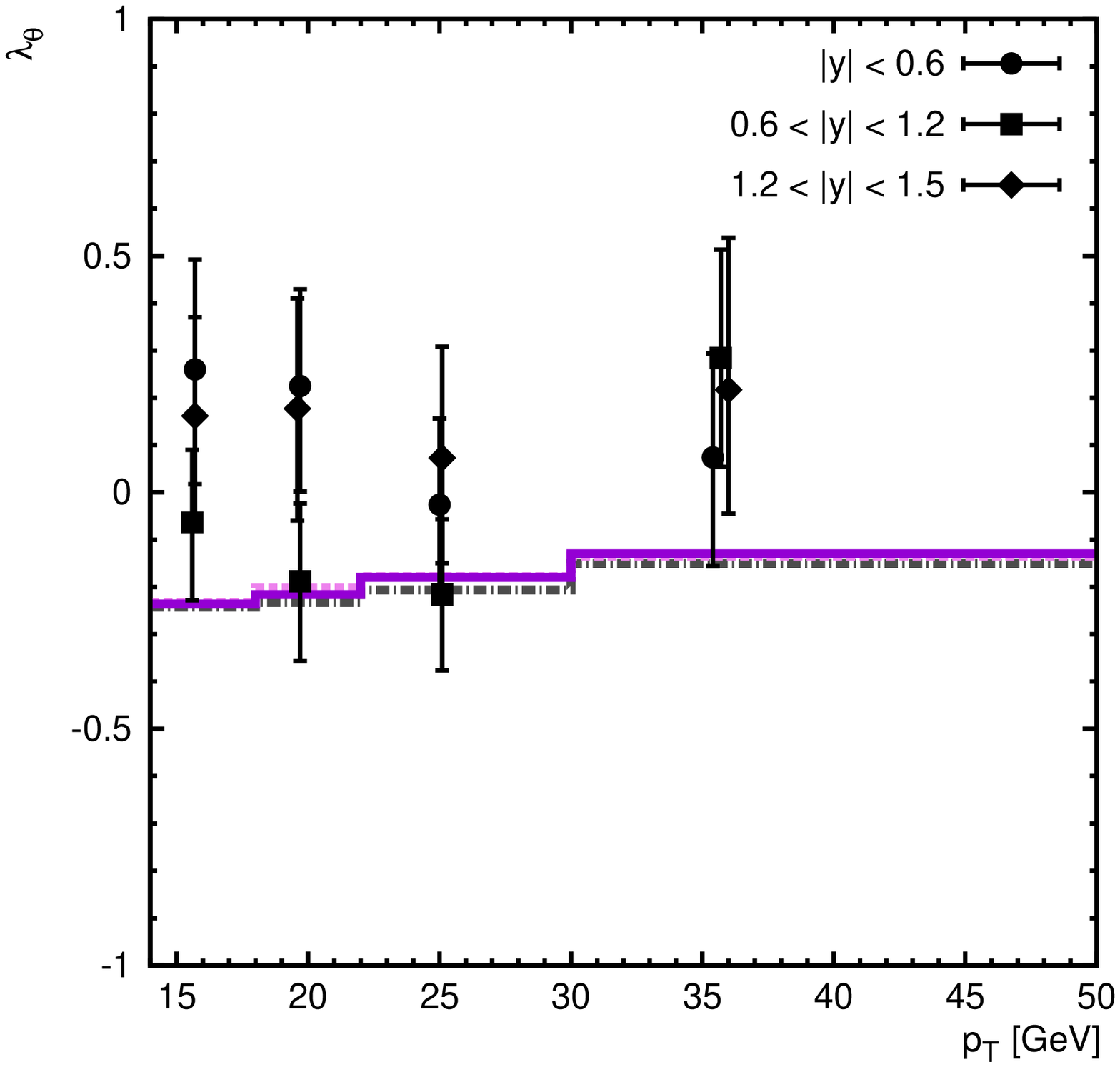, width = 6cm, angle = 270}
\vspace{0.7cm} \hspace{-1cm}
\epsfig{figure=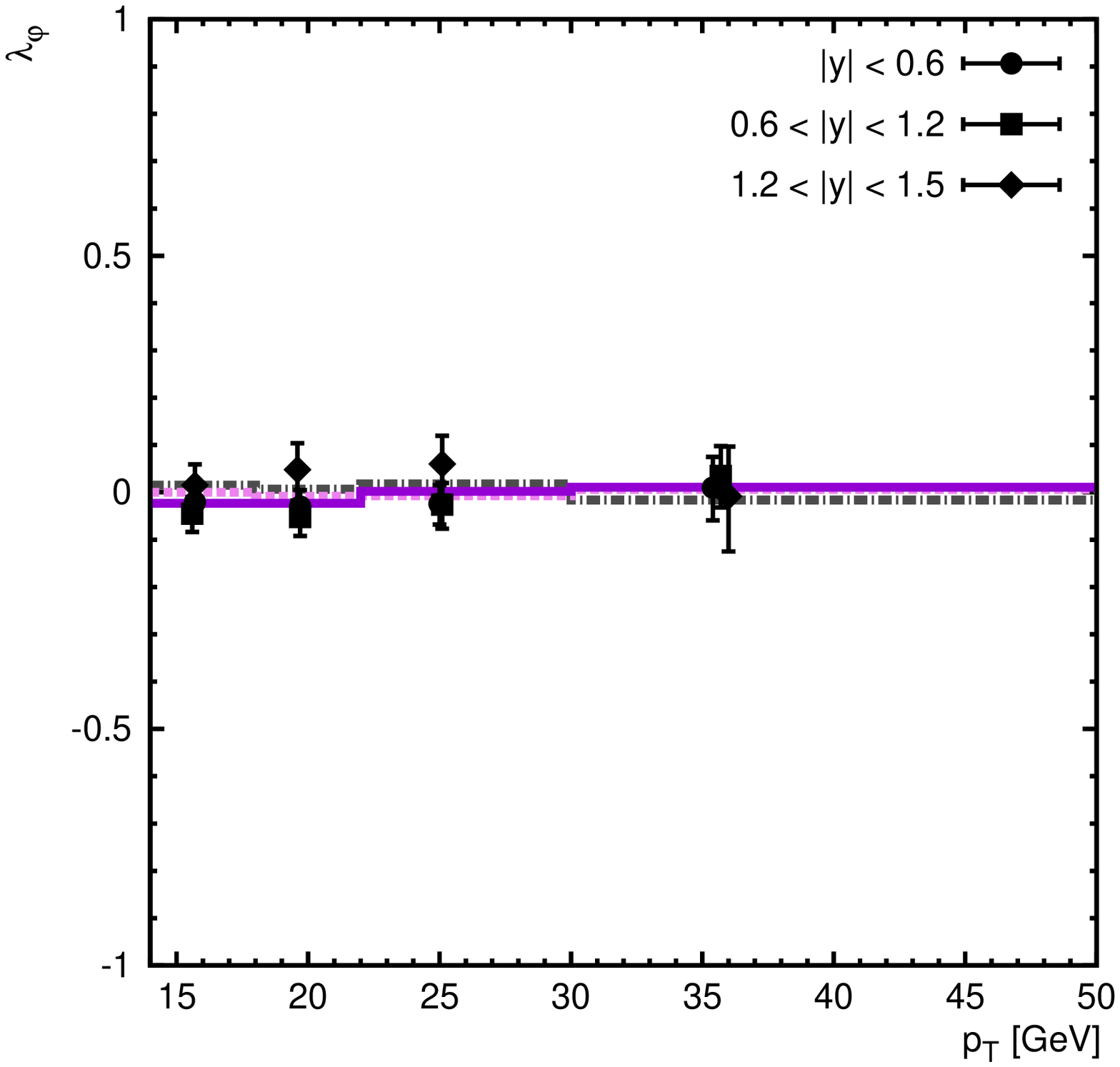, width = 6cm, angle = 270}
\epsfig{figure=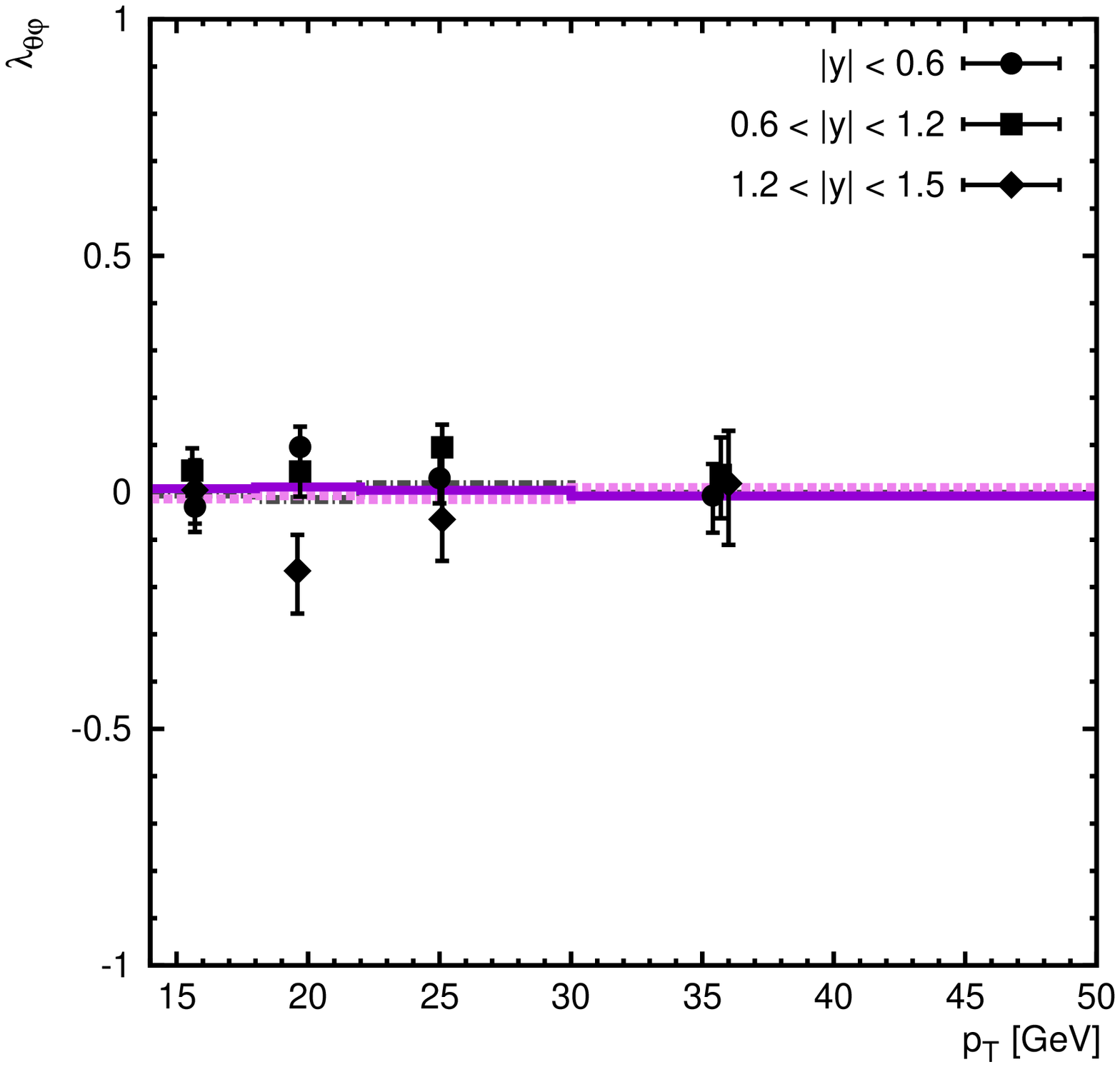, width = 6cm, angle = 270}
\hspace{-1cm} 
\epsfig{figure=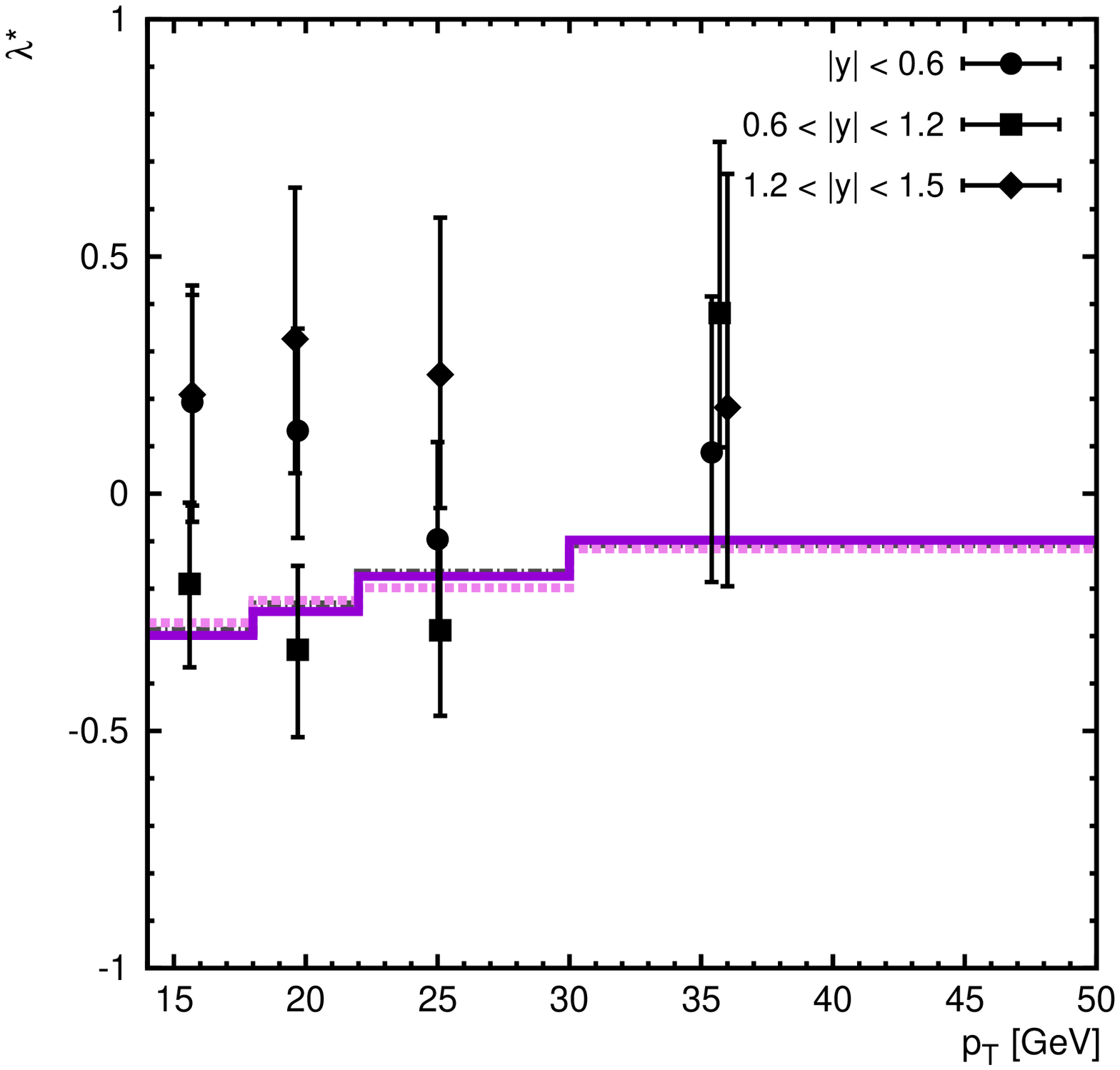, width = 6cm, angle = 270}
\caption{Polarization parameters 
$\lambda_\theta$, $\lambda_\phi$, $\lambda_{\theta \phi}$
and $\lambda^*$ of prompt $\psi(2S)$ mesons 
calculated as a function of $\psi(2S)$ transverse momentum in the helicity frame.
The solid, dashed and dash-dotted histograms correspond
to the predictions obtained at $|y| < 0.6$, $0.6 < |y| < 1.2$
and $1.2 < |y| < 1.5$. The KMR gluon distribution is used.
The experimental data are from CMS\cite{20}.}
\label{fig5}
\end{center}
\end{figure}

\begin{figure}
\begin{center}
\epsfig{figure=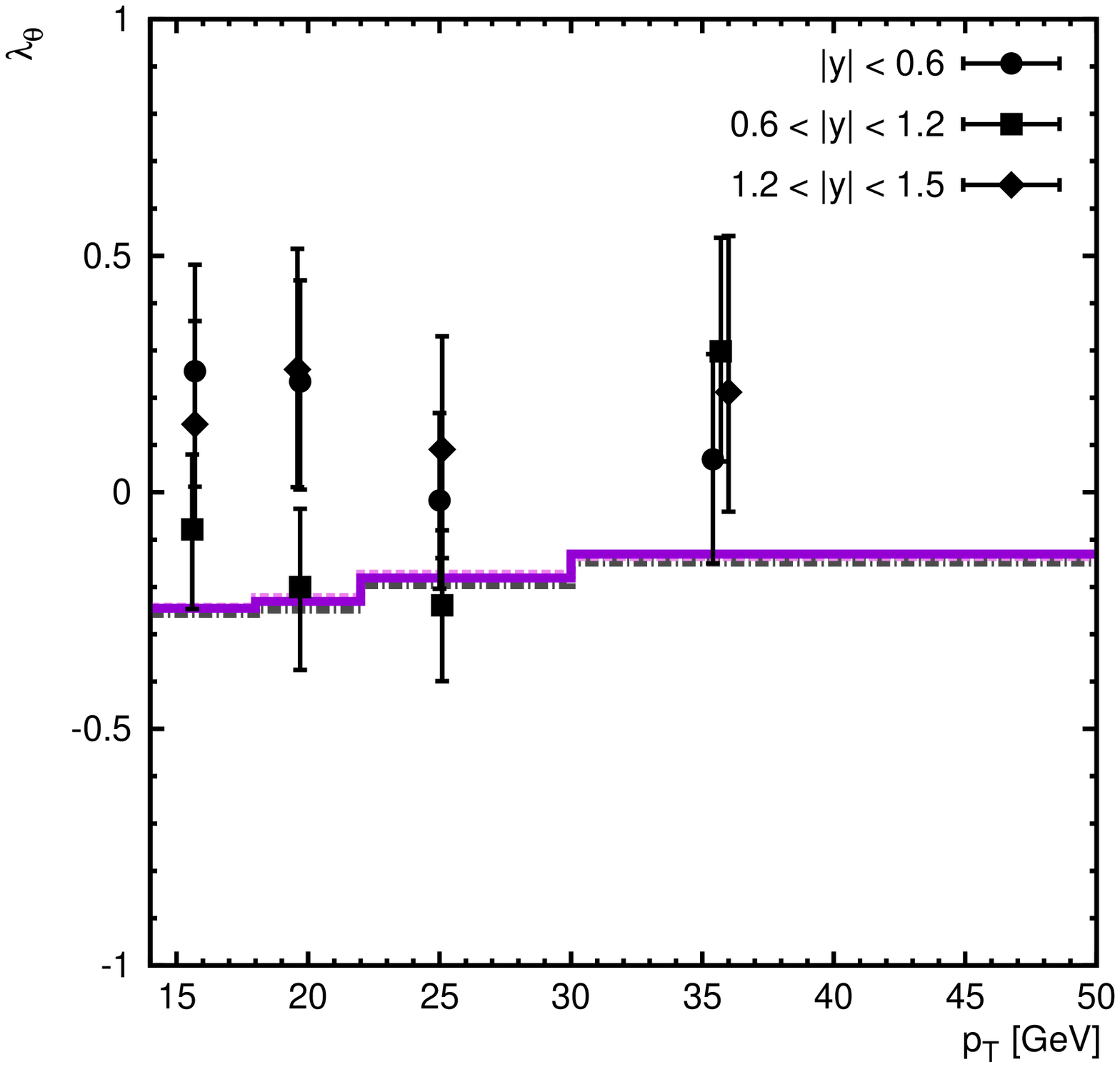, width = 6cm, angle = 270}
\vspace{0.7cm} \hspace{-1cm}
\epsfig{figure=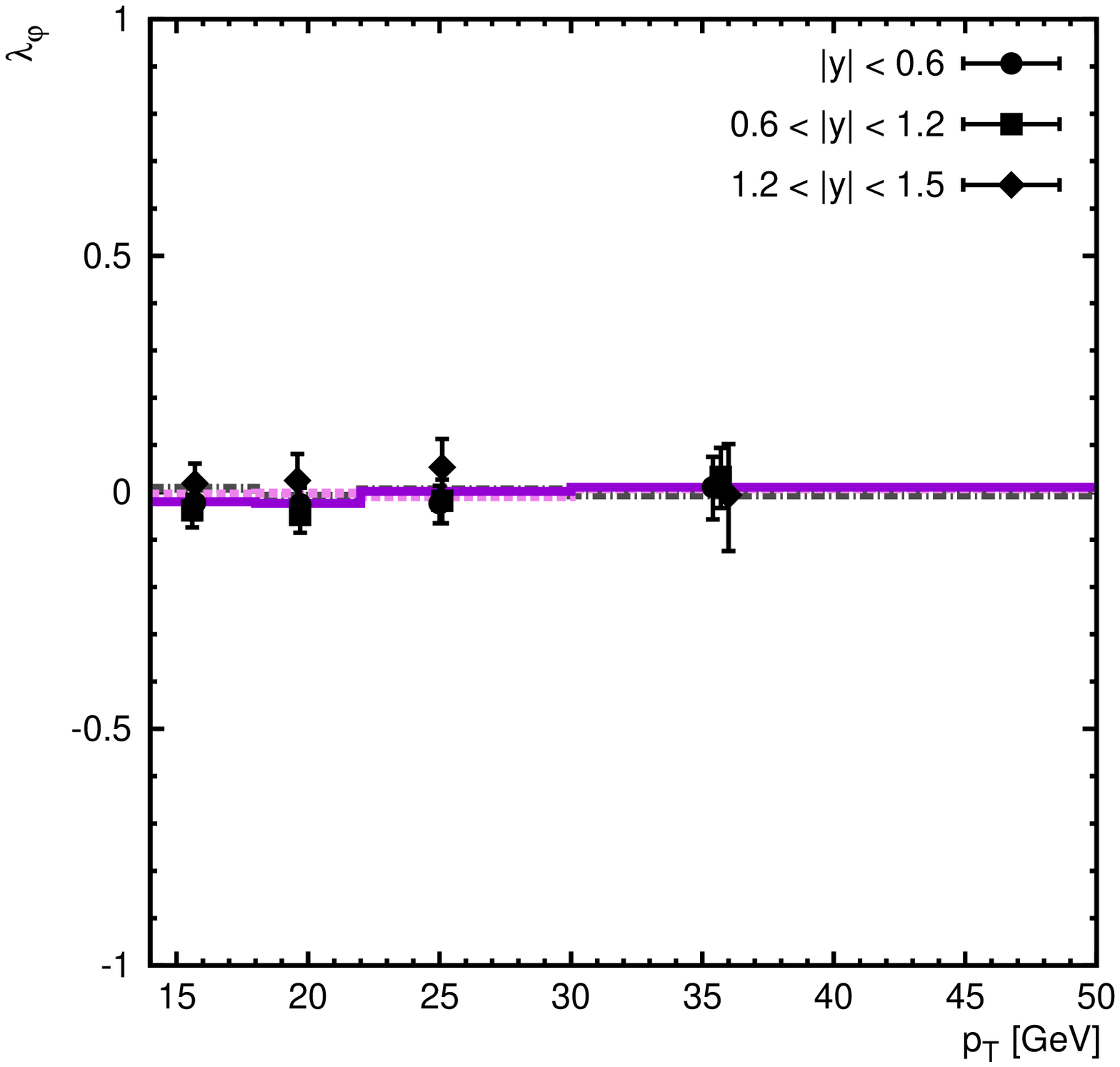, width = 6cm, angle = 270}
\epsfig{figure=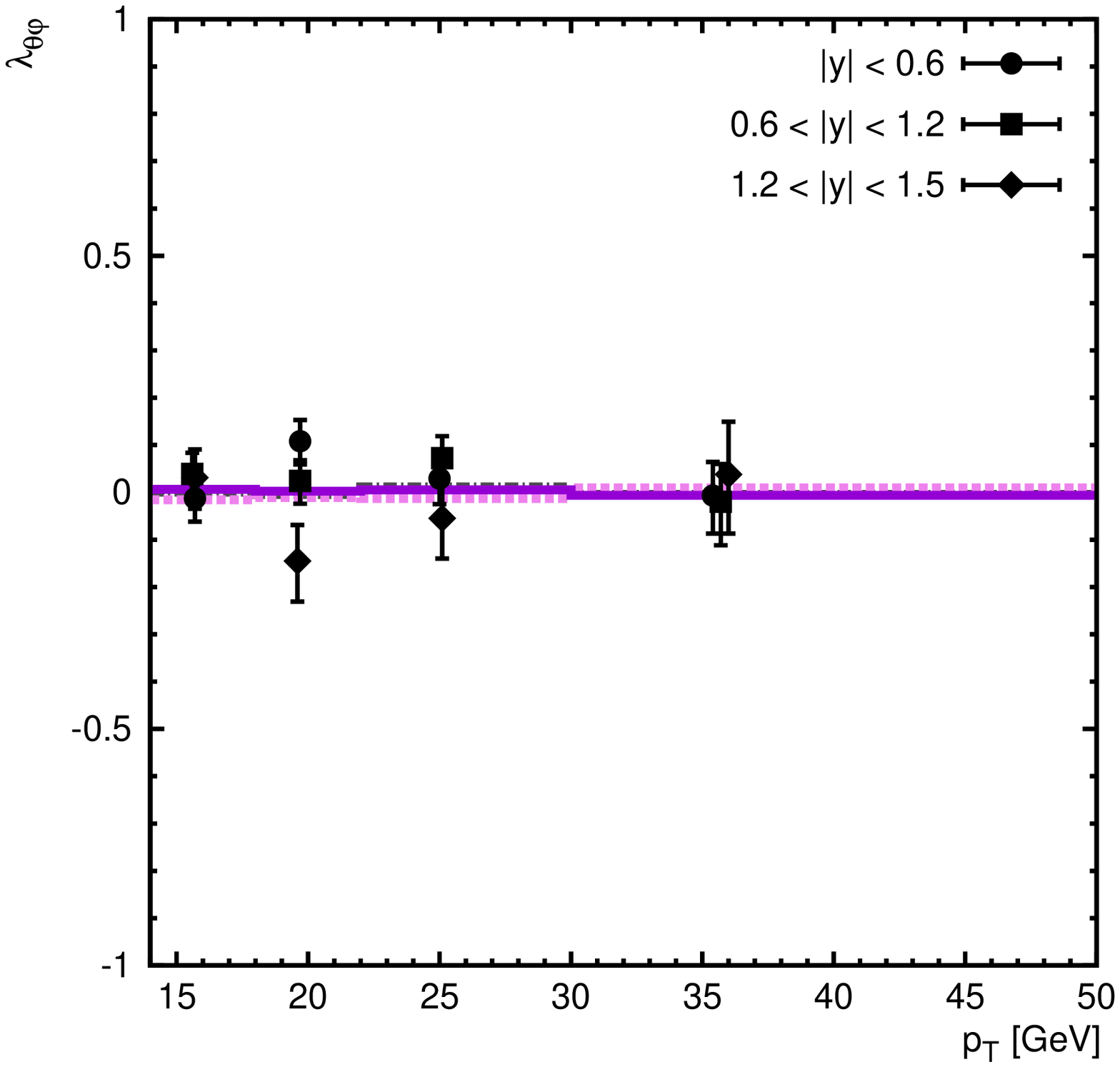, width = 6cm, angle = 270}
\hspace{-1cm} 
\epsfig{figure=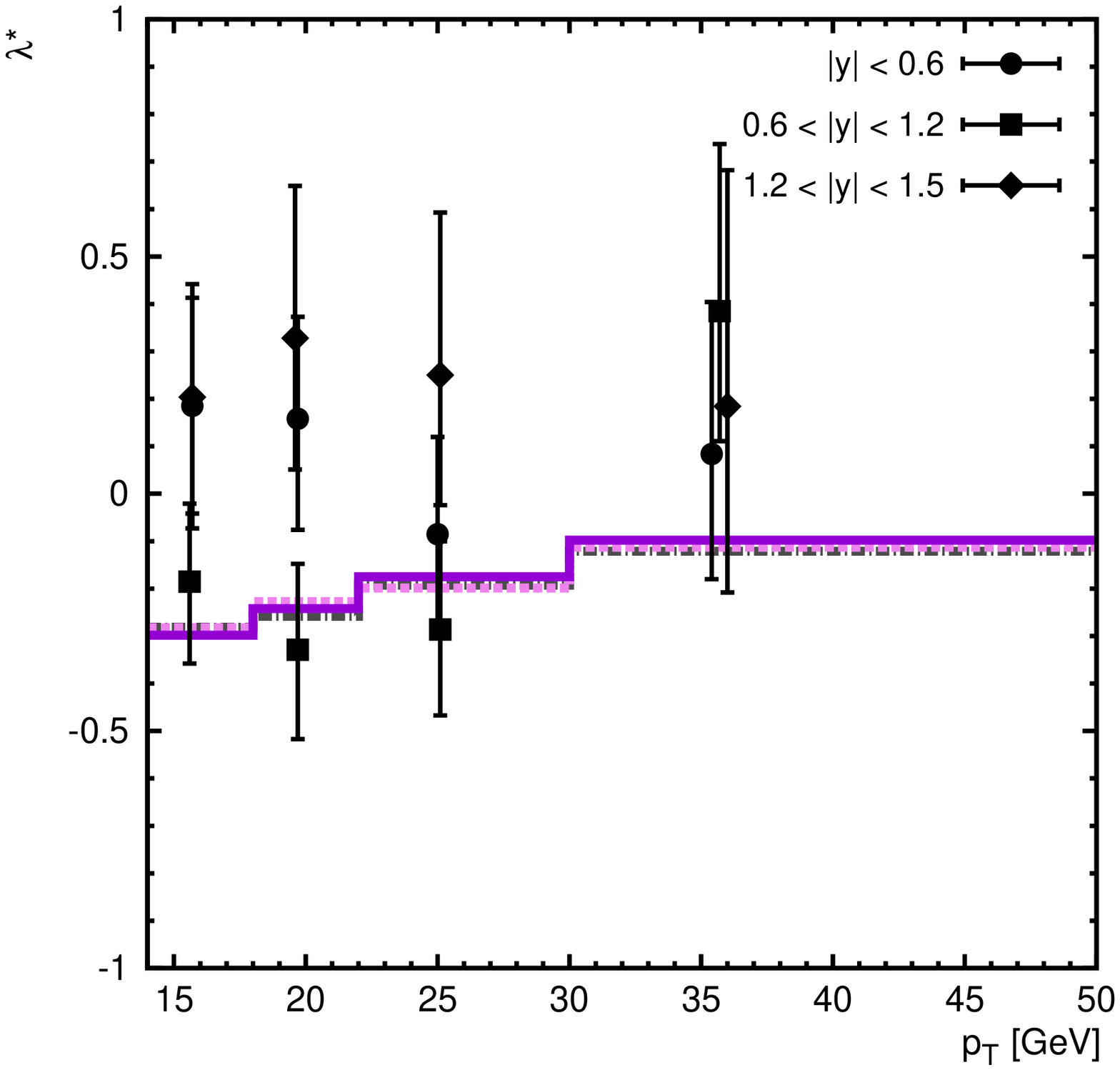, width = 6cm, angle = 270}
\caption{Polarization parameters 
$\lambda_\theta$, $\lambda_\phi$, $\lambda_{\theta \phi}$
and $\lambda^*$ of prompt $\psi(2S)$ mesons 
calculated as a function of $\psi(2S)$ transverse momentum in the perpendicular helicity frame.
The solid, dashed and dash-dotted histograms correspond
to the predictions obtained at $|y| < 0.6$, $0.6 < |y| < 1.2$
and $1.2 < |y| < 1.5$. The KMR gluon distribution is used.
The experimental data are from CMS\cite{20}.}
\label{fig6}
\end{center}
\end{figure}

\begin{figure}
\begin{center}
\epsfig{figure=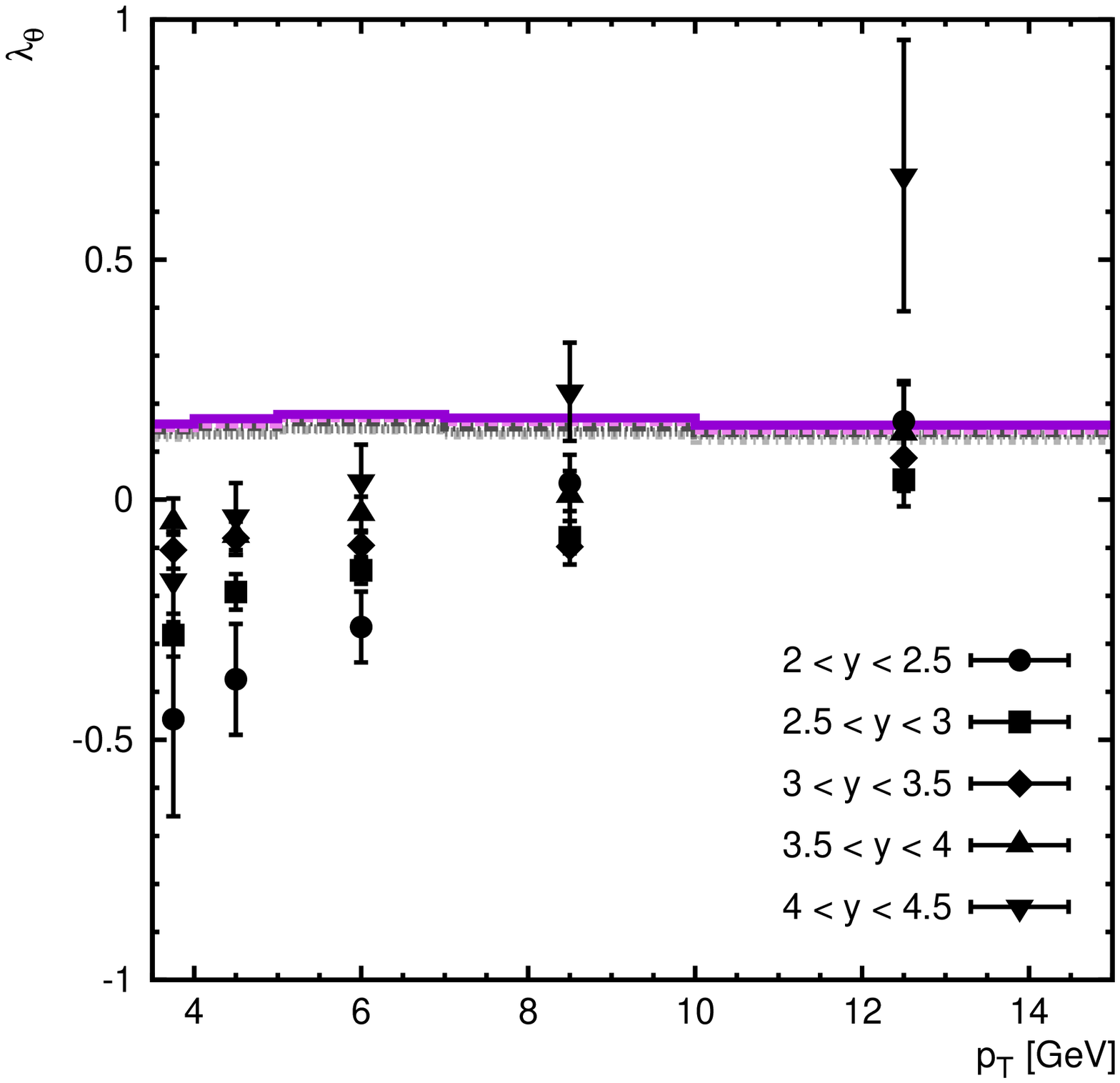, width = 6cm, angle = 270}
\vspace{0.7cm} \hspace{-1cm}
\epsfig{figure=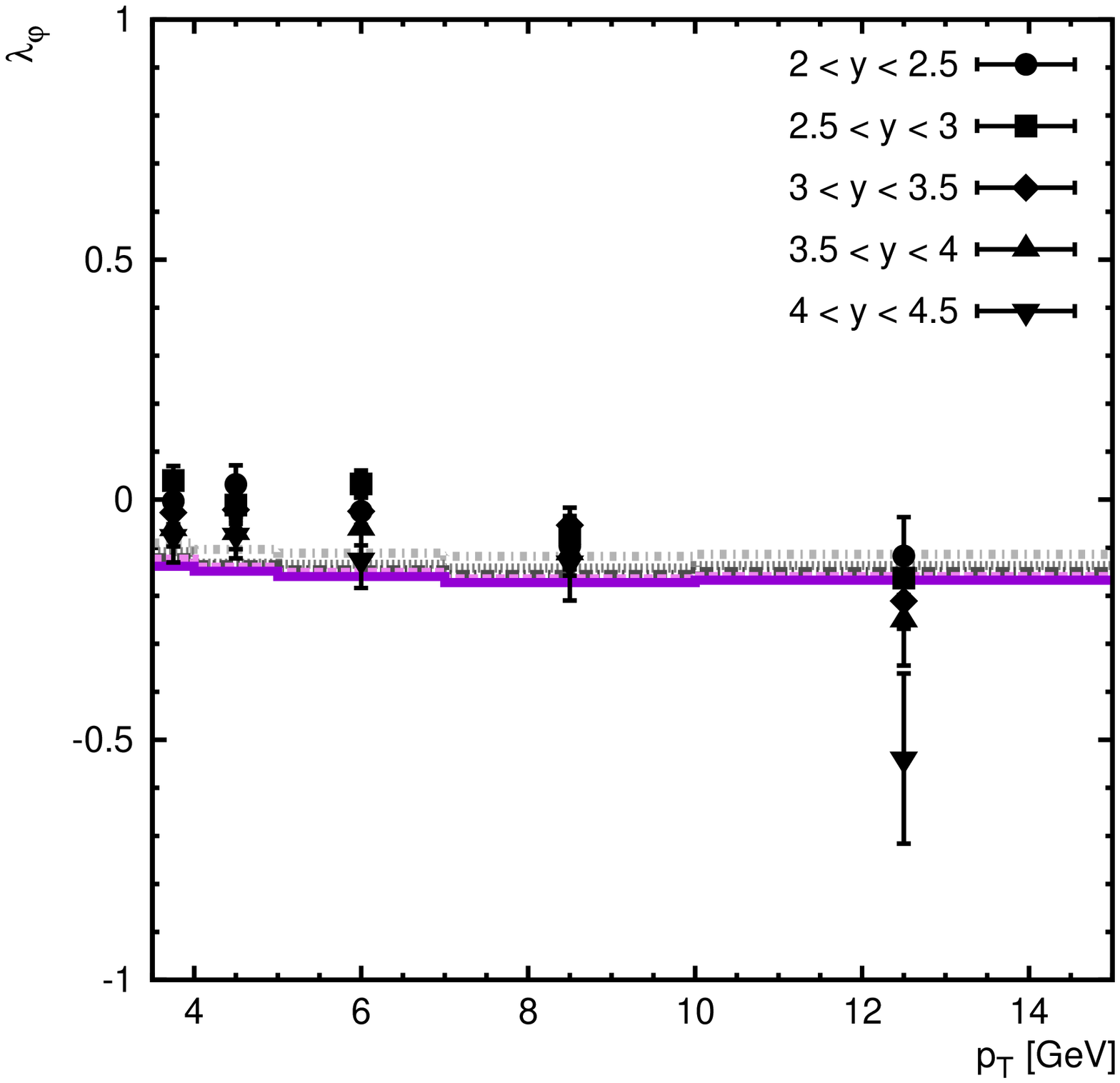, width = 6cm, angle = 270}
\epsfig{figure=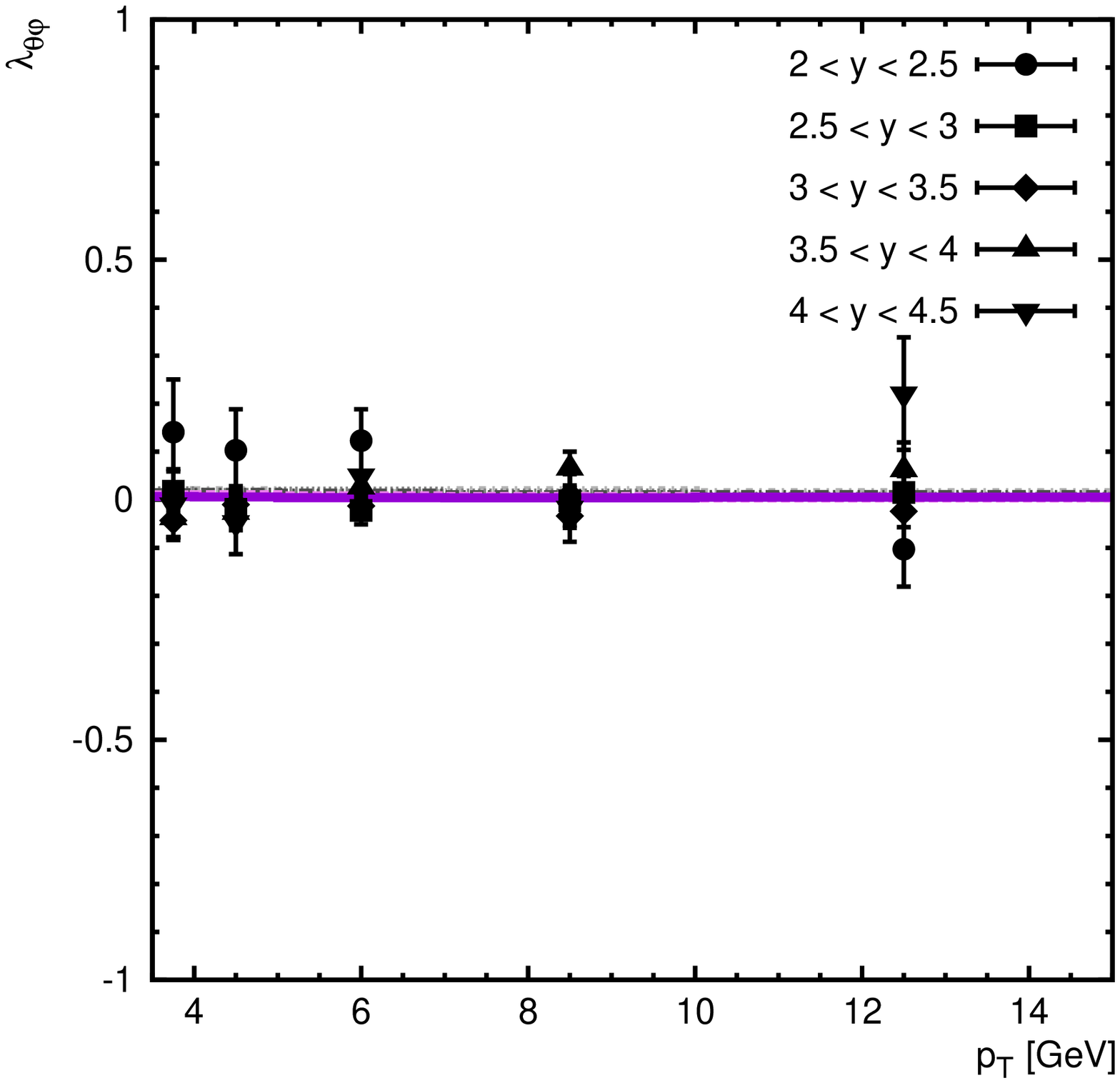, width = 6cm, angle = 270}
\hspace{-1cm} 
\epsfig{figure=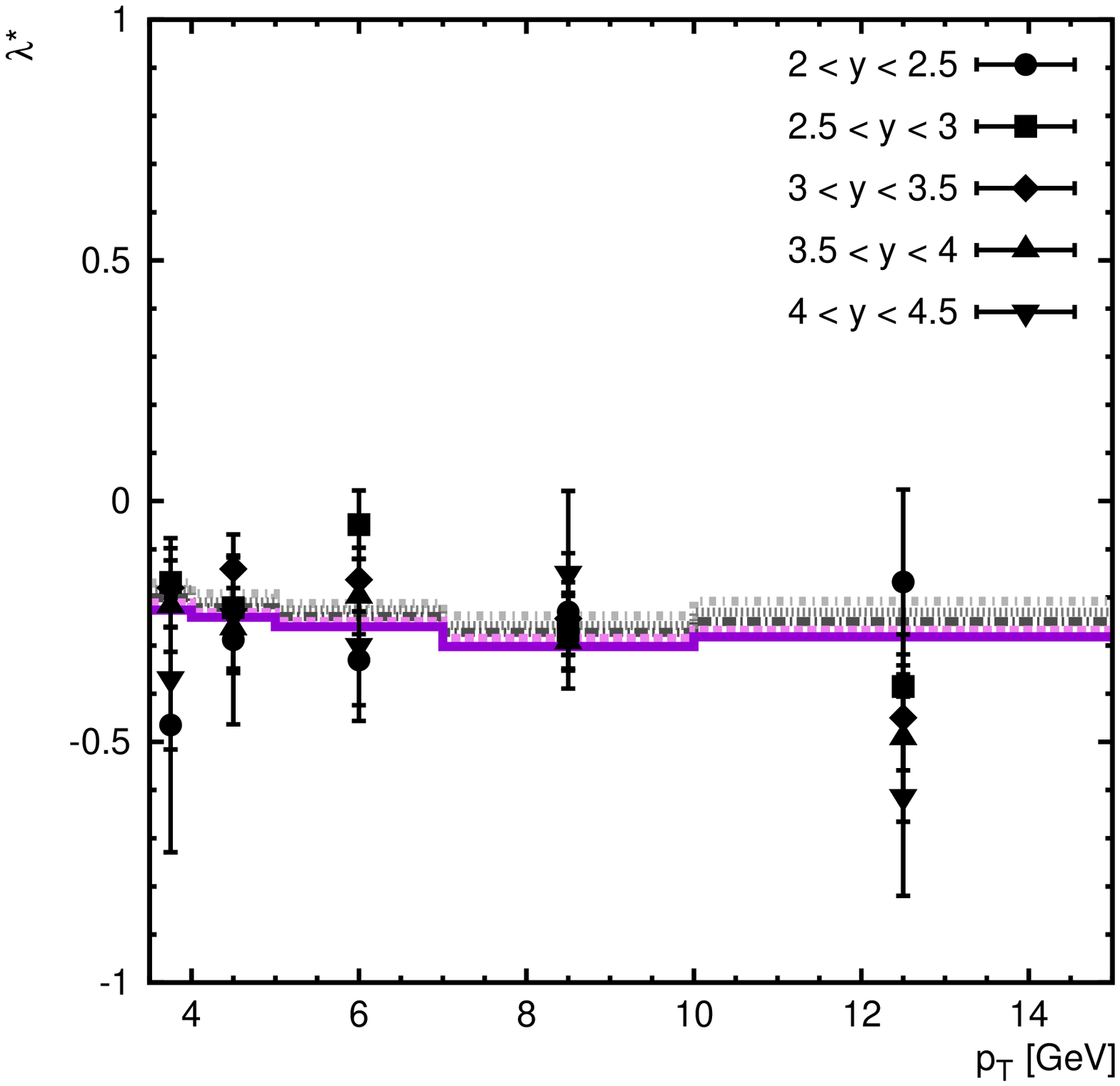, width = 6cm, angle = 270}
\caption{Polarization parameters 
$\lambda_\theta$, $\lambda_\phi$, $\lambda_{\theta \phi}$
and $\lambda^*$ of prompt $\psi(2S)$ mesons 
calculated as a function of $\psi(2S)$ transverse momentum in the Collins-Soper frame.
The solid, dashed, dash-dotted, dotted and short dash-dotted histograms correspond
to the predictions obtained at $2 < y < 2.5$, $2.5 < y < 3$,
$3 < y < 3.5$, $3.5 < y < 4$ and $4 < y < 4.5$. 
The KMR gluon distribution is used.
The experimental data are from LHCb\cite{21}.}
\label{fig7}
\end{center}
\end{figure}

\begin{figure}
\begin{center}
\epsfig{figure=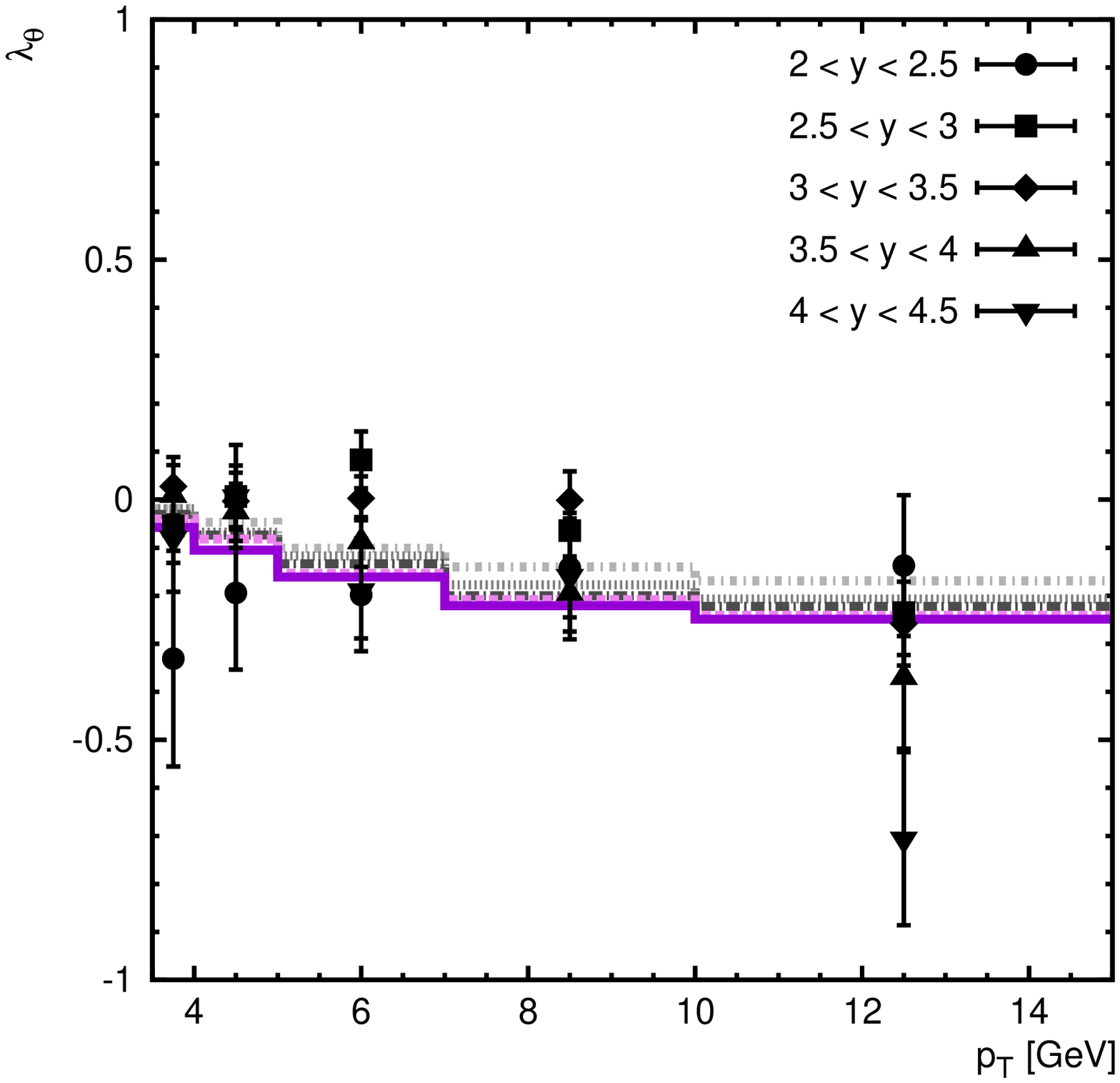, width = 6cm, angle = 270}
\vspace{0.7cm} \hspace{-1cm}
\epsfig{figure=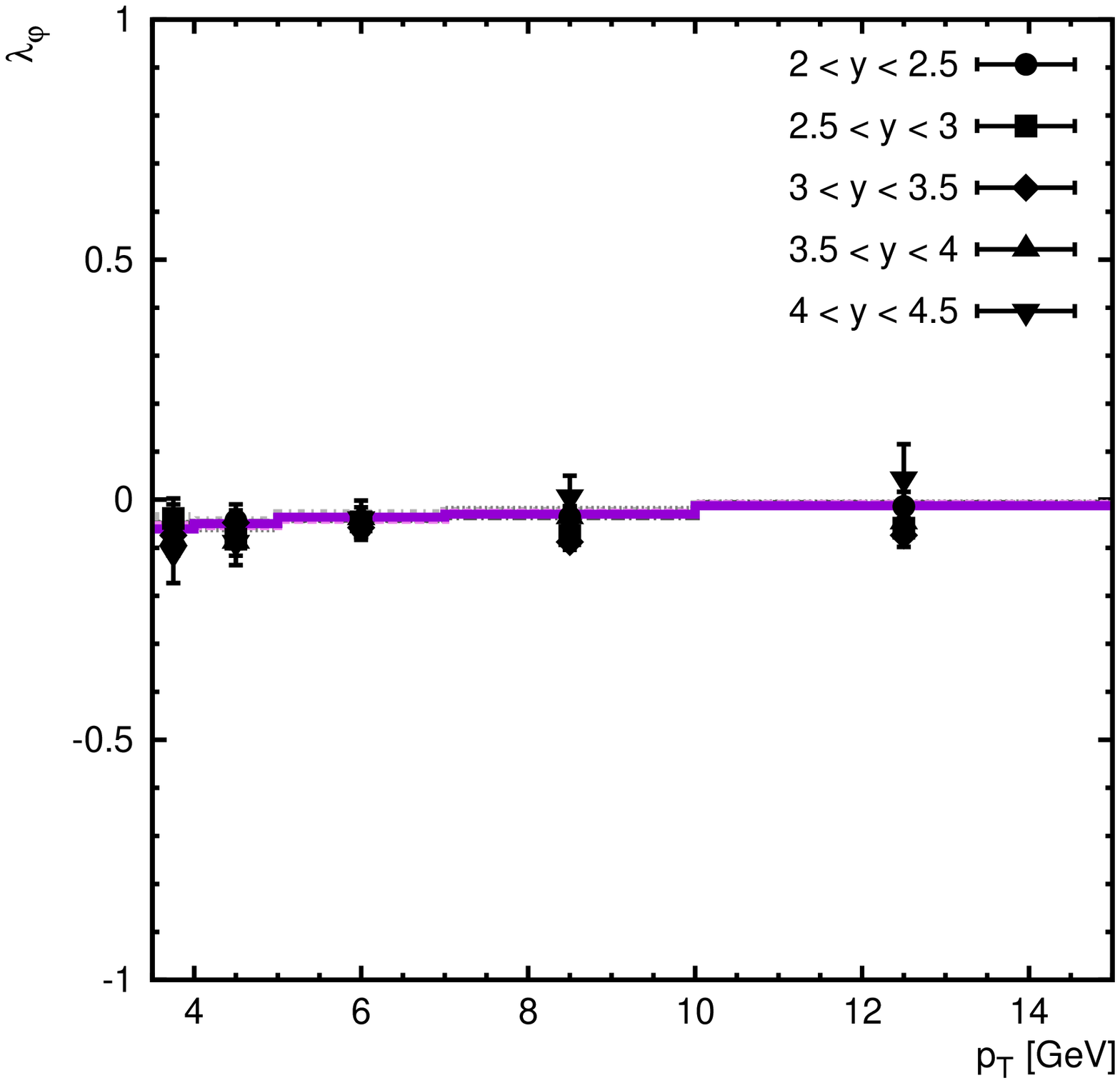, width = 6cm, angle = 270}
\epsfig{figure=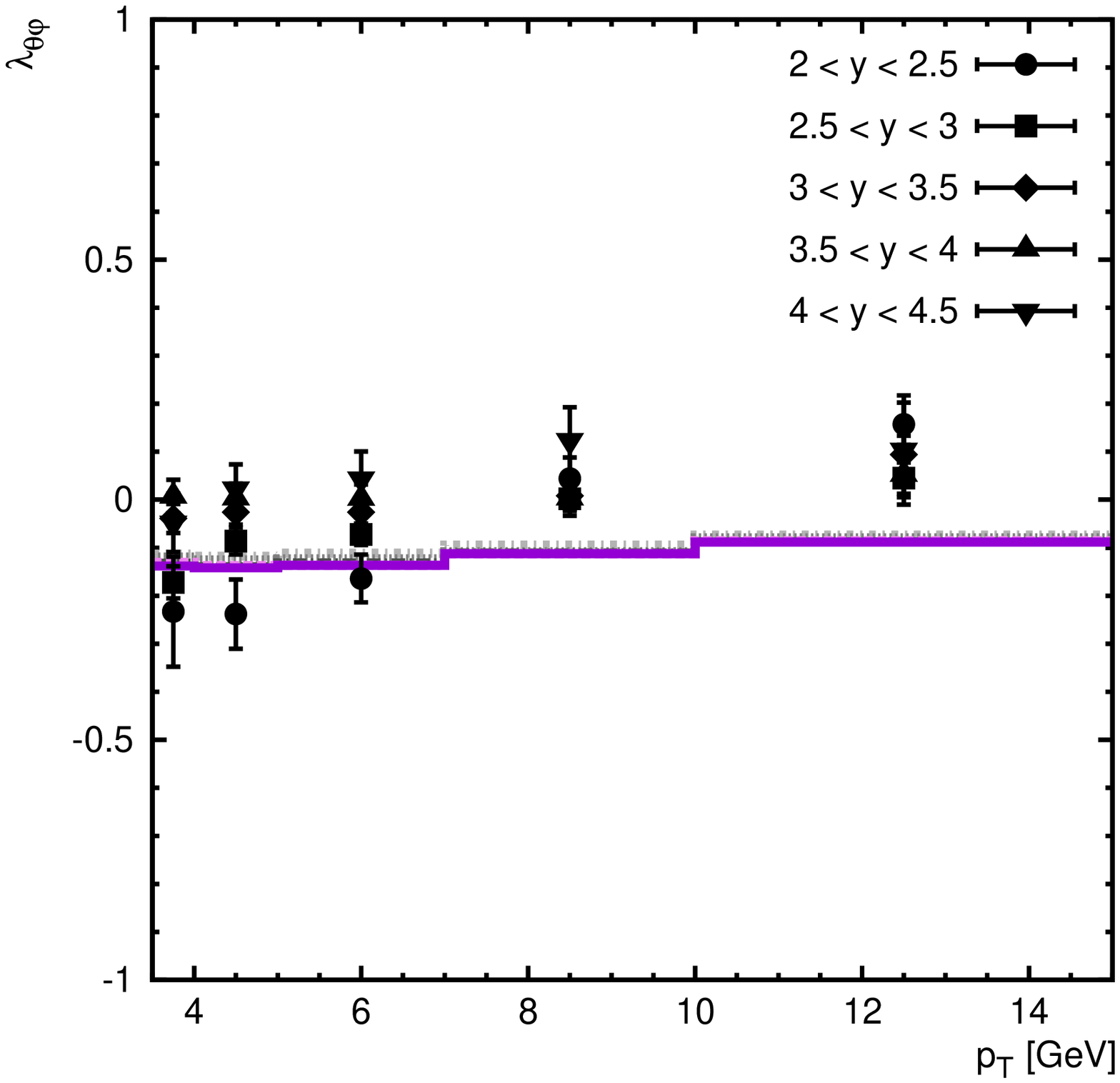, width = 6cm, angle = 270}
\hspace{-1cm} 
\epsfig{figure=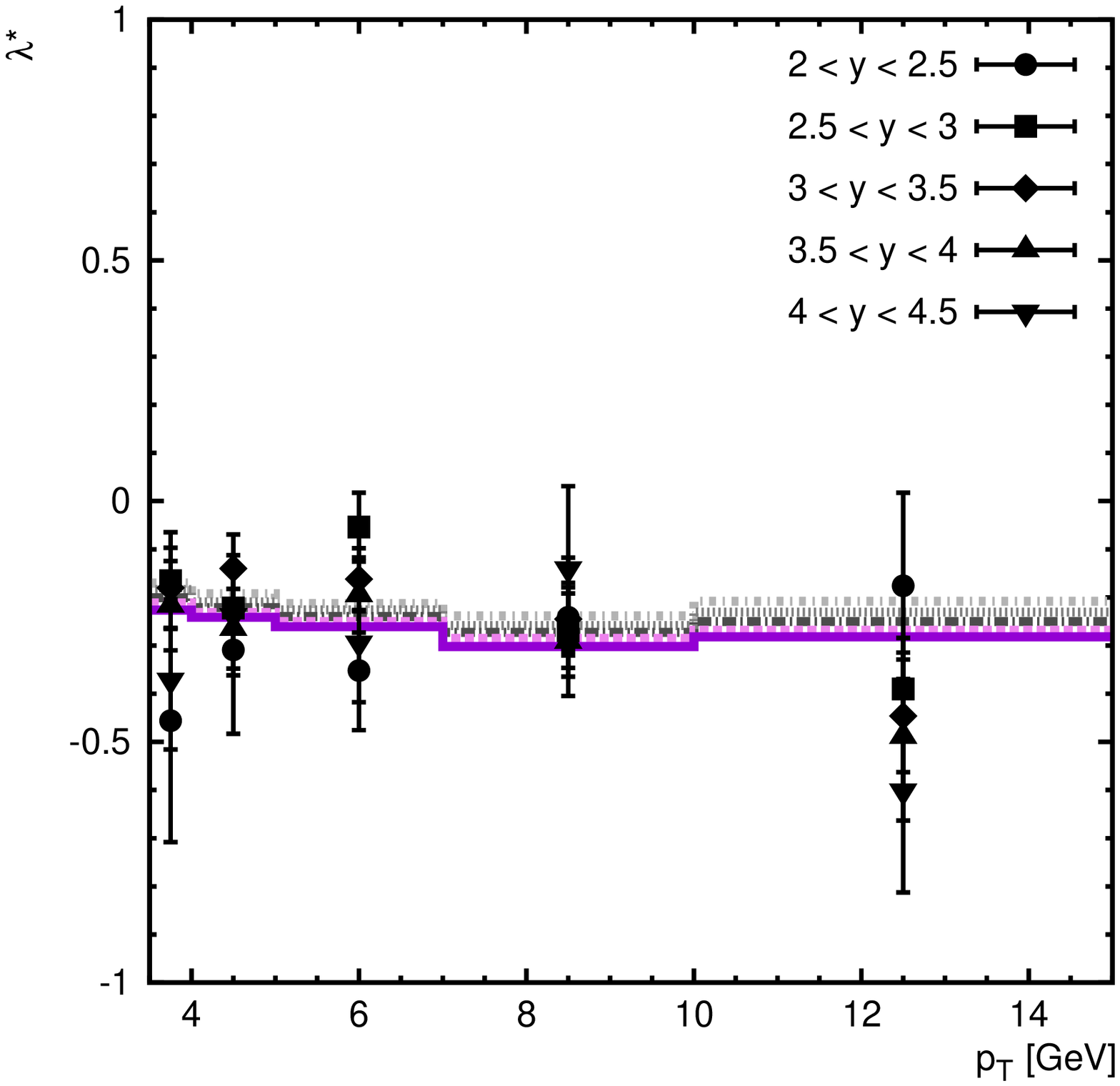, width = 6cm, angle = 270}
\caption{Polarization parameters 
$\lambda_\theta$, $\lambda_\phi$, $\lambda_{\theta \phi}$
and $\lambda^*$ of prompt $\psi(2S)$ mesons 
calculated as a function of $\psi(2S)$ transverse momentum in the helicity frame.
The solid, dashed, dash-dotted, dotted and short dash-dotted histograms correspond
to the predictions obtained at $2 < y < 2.5$, $2.5 < y < 3$,
$3 < y < 3.5$, $3.5 < y < 4$ and $4 < y < 4.5$. 
The KMR gluon distribution is used.
The experimental data are from LHCb\cite{21}.}
\label{fig8}
\end{center}
\end{figure}

\end{document}